 \documentclass[prc,showpacs,preprintnumbers,twocolumn,
                superscriptaddress,amsmath,amssymb,floatfix]{revtex4}
\usepackage{dcolumn}
\usepackage{bm}
\usepackage{longtable}
\usepackage{dsfont}

 \usepackage{graphicx,epsfig,latexsym,amssymb}
 \usepackage{multirow,amsmath,array,booktabs,color}
 \usepackage[section]{placeins}

\newcommand{\beq}{\vspace{0.5em}\begin{equation}}
\newcommand{\eeq}{\end{equation}\vspace{0.5em}}
\newcommand{\beqn}{\vspace{0.5em}\begin{eqnarray}}
\newcommand{\eeqn}{\end{eqnarray}\par\vspace{0.5em}\noindent}

\newcommand{\beqa}{\vspace{0.5em}\begin{eqnarray*}}
\newcommand{\eeqa}{\end{eqnarray*}\par\vspace{0.5em}}
\newcommand{\bea}{\begin{array}}
\newcommand{\eea}{\end{array}}

\newcommand{\bcen}{\begin{center}}
\newcommand{\ecen}{\end{center}}
\newcommand{\btab}{\begin{tabular}}
\newcommand{\etab}{\end{tabular}}
\newcommand{\bsub}{\begin{subequations}}
\newcommand{\esub}{\end{subequations}}

\newcommand{\kap}{\kappa}

\newcommand{\bsig}{\mbox{\boldmath$\sigma$}}

\newcommand{\btau}{\mbox{\boldmath$\tau$}}

\newcommand{\bj}{{\mathbf{j}}}

\newcommand{\bp}{{\mathbf{p}}}

\newcommand{\bP}{{\mathbf{P}}}

\newcommand{\bV}{{\mathbf{V}}}

\newcommand{\dfra}{\displaystyle\frac}
\begin{document}

 \title{Three-dimensional angular momentum projection in relativistic mean-field theory}
 \author{J. M. Yao}
  \email{jmyao@pku.edu.cn}
 \address{State Key Lab Nucl. Phys. $\&$ Tech., School of Physics, Peking University,
         Beijing 100871, China}
 \address{Physik-Department der Technischen Universit\"at M\"unchen, D-85748
 Garching, Germany}
 \author{J. Meng}
  \email{mengj@pku.edu.cn}
 \address{State Key Lab Nucl. Phys. $\&$ Tech., School of Physics,
          Peking University, Beijing 100871, China}
 \address{Institute of Theoretical Physics, Chinese Academy of
          Sciences, Beijing, China}
 \address{Center of Theoretical Nuclear Physics, National
          Laboratory of Heavy Ion Accelerator, 730000 Lanzhou, China}
 \author{P. Ring}
 \email{ring@ph.tum.de}
 \address{Physik-Department der Technischen Universit\"at M\"unchen, D-85748
         Garching, Germany}
 \author{D.~Pena~Arteaga }
 \address{Physik-Department der Technischen Universit\"at M\"unchen, D-85748
         Garching, Germany}
\date{\today}
 \vspace{2em}
 \begin{abstract}%
Based on a relativistic mean-field theory with an effective point
coupling between the nucleons, three-dimensional angular momentum
projection is implemented for the first time to project out states
with designed angular momentum from deformed intrinsic states
generated by triaxial quadrupole constraints.  The same effective
parameter set PC-F1 of the effective interaction is used for deriving
the mean field and  the collective Hamiltonian. Pairing correlations
are taken into account by the BCS method using both monopole forces
and zero range $\delta$-forces with strength parameters adjusted to
experimental even-odd mass differences. The method is applied
successfully to the isotopes $^{24}$Mg, $^{30}$Mg, and $^{32}$Mg.
 \end{abstract}
 \maketitle

 \section{Introduction}

\label{Sec.I}

Experimental and theoretical studies of nuclei far from the
$\beta$-stability line are at the forefront of nuclear science. Until
1985~\cite{Tanihata85HI}, the access to nuclei near the border of
$\beta$-stability was practically impossible. The advent of
radioactive ion beams (RIBs)~\cite{Tanihata85PRL,Bertulani01}
provides a useful tool for studying the structure of such unstable
nuclei. Hitherto, RIBs have already disclosed many structure
phenomena in exotic nuclei with extreme isospin values, and the next
generation of radioactive-beam facilities will present new exciting
opportunities for the study of the nuclear many-body
systems~\cite{Mueller93,Tanihata95,Hansen95,Casten00,Mueller01,Jonson04,Jensen04}.

Energy density functional (EDF) theory in nuclear physics is nowadays
the most important microscopic approach for large-scale nuclear
structure calculations in heavy nuclei and it has been successfully
employed for the description of nuclei far from
$\beta$-stability~\cite{Bender03,Vretenar05}. The nuclear EDF is
constructed phenomenologically, based on the knowledge accumulated
within modern self-consistent mean-field (SCMF) approaches built upon
an effective density-dependent two-body interaction. Compared with
the shell model approach~\cite{Otsuka01,CMZ.05}, EDF functionals are
universal in the sense that they can be applied to nuclei all over
the periodic table. Because of its simplicity SCMF approaches have a
great advantage in particular for the description of heavy exotic
nuclei.

The great success achieved by SCMF theories in the description of
nuclear properties relies on the fact that within these theories the
complicated many-body wave functions are approximated by a single
Slater determinant. Important many-body correlations are taken into
account via the mechanism of \textquotedblleft spontaneous symmetry
breaking\textquotedblright~\cite{Ring80}. Examples are the violation
of SO(3) rotational symmetry in deformed nuclei and of U(1) symmetry
in gauge space in superfluid nuclei. As a consequence, such product
wave functions are not eigenstates of the angular momentum and
particle number operators. These deficiencies give rise to several
serious problems in the description of particular nuclear properties,
as the absence of correlations associated with the symmetry
restoration, the admixture of low-lying excited states into the
ground state, difficulties in the connection to the laboratory frame
for spectroscopic observables, the absence of selection rules for
transitions, etc. Therefore, in order to compare properly with the
experimental data, one has to go beyond the mean-field approximation.
Projection methods provide an effective tool to restore the
spontaneous breaking of
symmetries~\cite{Yoc.57,PY.57,Zeh.65,MacDonald70,Wong75}. A suitable
linear combination of intrinsic states deformed in Euler space or
gauge space recover rotational or gauge symmetry. Such procedures are
known as Angular Momentum Projection (AMP) or Particle Number
Projection (PNP) methods.

Angular Momentum Projection has been a goal of nuclear physicists for
many years. However, due to its numerical complexity, only in the
last ten years it has been possible to apply such projection
procedure in the context of SCMF theory with realistic effective
forces, for example the non-relativistic Skyrme force
SLy4~\cite{Valor00}, the Gogny force D1S~\cite{RER.02a,Guzman02npa}
or the relativistic point coupling force
PC-F1~\cite{Niksic06I,Niksic06II}. These investigations have shown
that the energy gain due to the restoration of rotational symmetry is
of the order of several MeV and it has great influence on the
topological structure of the nuclear potential energy surface (PES).
In these three cases, however, axial symmetry in the mean-fields has
been imposed from the beginning. Such a restriction simplifies the
numerical problem considerably, because in this case, the integrals
over two of the three Euler angles in the kernels can be treated
analytically and one is left with a one-dimensional integration.

As illustrated by recent systematic calculations~\cite{Moller06},
specific combinations of single-particle orbitals near the Fermi
surface and the additional binding energy due to non-axial degrees of
freedom can enhance the tendency to form nuclei with triaxial shapes.
Several \textit{islands of triaxiality} have been revealed throughout
the nuclear chart. The inclusion of triaxiality can dramatically
reduce the barrier separating prolate and oblate minima, leading to
structures that are soft or unstable for triaxial
distortions~\cite{Cwiok05}. Furthermore, the occurrence of
triaxiality can give rise to many very interesting modes of
collective motion, which are very different from those of axially
deformed shape, such as Chiral rotation~\cite{Grodner06}, Wobbling
motion~\cite{Odegard01} and the violation of $K$-selection rules in
electromagnetic transitions~\cite{Chowdhury88}.

To describe properly the properties of possible triaxially deformed
nuclei and especially to examine the role of triaxial deformation in
the context of SCMF theory, it is essential to introduce the
$\gamma$-degree of freedom at the mean-field level and to perform
full three-dimensional angular momentum projection (3DAMP). In the
context of phenomenological models with small shell model spaces and
the corresponding effective interactions, 3DAMP has already been
implemented many years ago in
Refs.~\cite{Giraud69,Hara82,Hayashi84,Burzynski95,Enami99}. The
restoration of rotational symmetry has been shown to have a strong
influence on the topological structure of the ($\beta, \gamma$)
energy surface for transitional nuclei~\cite{Hayashi84}. In
particular, the correlations taken into account by 3DAMP are found to
have a tendency to lower the potential energy in the region of strong
triaxial deformations~\cite{Enami00ptp}. In the context of energy
density functionals, 3DAMP has been performed on top of Hartree-Fock
(HF) with a simple Skyrme-type interaction~\cite{Baye84}, or with the
full Skyrme energy functional~\cite{Zdunczuk07}. In both cases
cranked wave functions were projected to approximate a variation
after projection procedure, but pairing correlations were not
included. Only very recently, 3DAMP+PNP with configuration mixing has
been attempted in the context of triaxial Hartree-Fock-Bogoliubov
(HFB) theory with the full Skyrme energy functional~\cite{Bender08}.

During the past decades, relativistic mean-field (RMF) theory, which
relies on basic ideas of effective field theory~\cite{LNP.641} and
of density functional theory~\cite{KS.65} has achieved great success
in describing many nuclear phenomena for both stable and exotic
nuclei over the entire nuclear chart with a few universal
parameters~\cite{Serot86,Reinhard89,Ring96,Vretenar05,Meng06}. It
incorporates many important relativistic effects, such as the
presence of large Lorentz scalar and vector fields with
approximately equal magnitude and opposite sign. This leads to a new
saturation mechanism via the difference between the scalar and
vector densities, and naturally to the large spin-orbit splitting
needed for the understanding of magic numbers in finite nuclei.
Moreover, relativistic effects are responsible for the efficient
description of spin observables in medium-energy proton-nucleus
scattering using the relativistic impulse
approximation~\cite{Murdock86} and for the existence of approximate
pseudospin symmetry in nuclear spectra~\cite{Arima69,Ginocchio97}.
All these features motivate further investigations in the framework
of RMF theory and new efforts to improve its predictive power.

The extension of RMF theory for the description of triaxially
deformed nuclei was first done decades ago~\cite{Koepf88}. Later it
has been employed in many studies on the effect of $\gamma$
deformation on nuclear
properties~\cite{Hirata96,Bender98,Meng06prc,Yao08}.
$\gamma$-deformation plays also an important role in the mean field
description of rotating nuclei in the framework of the cranking
model~\cite{Koepf89}: the Coriolis operator violates axial symmetry
and leads to currents and time-odd components in the intrinsic
nuclear fields~\cite{Afanasjev00b}. All these applications of
triaxial RMF theory are done on the mean field level. A full 3DAMP
for such cases is still missing and strongly desired, especially for
the description of transitional nuclei. In this work, we apply for
the first time 3DAMP to restore rotational symmetry for triaxially
deformed intrinsic states in the framework of RMF theory based on
point coupling interactions.

The paper has been arranged as follows. In Sec.~\ref{Sec.II} we
present an outline of the relativistic point coupling model that
will be used to generate mean-field wave functions with triaxial
symmetry, and we discuss three-dimensional angular momentum
projection. The method is applied for several isotopes, $^{24}$Mg,
$^{30}$Mg, and $^{32}$Mg to check the numerical accuracy of the code
as well as to present several illustrative results in
Sec.~\ref{Sec.III}. Finally, a summary and a perspective is given in
Sec.~\ref{Sec.IV}. Formulae of 3DAMP, and details about the
calculations of contractions and overlaps in the relativistic case
are collected in the Appendix.

 \section{Framework}
 \label{Sec.II}

 \subsection{The relativistic mean-field theory with point coupling}

 A detailed description of RMF theory with point coupling that will be adopted to
 generate intrinsic wave functions can be found in Ref.~\cite{Burvenich02}. In order
 to present a self-contained description of our approach we will give here a short outline
 of the relativistic point coupling model used in our applications.

 The elementary building blocks of a RMF theory with point coupling vertices are
 \beq
  \label{Block}
  (\bar \psi{\cal O}\Gamma\psi),\quad {\cal O}\in\{1,\vec\tau \},\quad
  \Gamma\in\{1,\gamma_\mu,\gamma_5,\gamma_5\gamma_\mu,\sigma_{\mu\nu}\},
 \eeq
 where $\psi$ is the Dirac spinor field of nucleon, $\vec\tau $ is the isospin vector
 and $\Gamma$ is one of the $4\times4$ Dirac matrices. There are ten such building blocks
 characterized by their transformation characteristics in isospin and in Minkowski space.
 We adopt arrows to indicate vectors in isospin space and bold types for the space vectors.
 Greek indices $\mu$ and $\nu$ run over the Minkowski indices 0, 1, 2, 3.

 A general effective Lagrangian can be written as a power series in $\bar \psi{\cal O}\Gamma\psi$
 and their derivatives. In present work, we start with the following Lagrangian
 density:
 \beq
  \label{Lagrangian}
  {\cal L} = {\cal L}^{\rm free} + {\cal L}^{\rm 4f}
             + {\cal L}^{\rm hot} + {\cal L}^{\rm der}
             + {\cal L}^{\rm em},
 \eeq
 where the Lagrangian density for free nucleon reads
 \beq
 {\cal L}^{\rm free} = \bar\psi(i\gamma_\mu\partial^\mu-m)\psi.
 \eeq
 The four-fermion point coupling term is given by
 \beqn
 {\cal L}^{\rm 4f}
             &=&-\frac{1}{2}\alpha_S(\bar\psi\psi)(\bar\psi\psi)
                -\frac{1}{2}\alpha_{TS}(\bar\psi\vec\tau\psi)\cdot(\bar\psi\vec\tau\psi)\nonumber\\
             && -\frac{1}{2}\alpha_{V}(\bar\psi\gamma_\mu\psi)(\bar\psi\gamma^\mu\psi)\nonumber\\
             &&
                -\frac{1}{2}\alpha_{TV}(\bar\psi\vec\tau\gamma_\mu\psi)\cdot(\bar\psi\vec\tau\gamma^\mu\psi),
 \eeqn
 which contains scalar-isoscalar, scalar-isovector, vector-isoscalar
 and vector-isovector channels. The medium dependence of the effective
 interaction has been taken into account by the higher order interaction terms
 \beq
 {\cal L}^{\rm hot}
                 = -\frac{1}{3}\beta_S(\bar\psi\psi)^3-\frac{1}{4}\gamma_S(\bar\psi\psi)^4
                   -\frac{1}{4}\gamma_V[(\bar\psi\gamma_\mu\psi)(\bar\psi\gamma^\mu\psi)]^2,
\label{E5}
 \eeq
 As in the nonrelativistic Skyrme functional~\cite{VB.72} gradient terms are essential. They simulate
 to some extent the effect of finite range of the force:
 \beqn
 {\cal L}^{\rm der}
             &=&-\frac{1}{2}\delta_S\partial_\nu(\bar\psi\psi)\partial^\nu(\bar\psi\psi)
                -\frac{1}{2}\delta_{TS}\partial_\nu(\bar\psi\vec\tau\psi)\cdot\partial^\nu(\bar\psi\vec\tau\psi)\nonumber\\
             &&
             -\frac{1}{2}\delta_V\partial_\nu(\bar\psi\gamma_\mu\psi)\partial^\nu(\bar\psi\gamma^\mu\psi)\nonumber\\
             &&
             -\frac{1}{2}\delta_{TV}\partial_\nu(\bar\psi\vec\tau\gamma_\mu\psi)\cdot\partial^\nu(\bar\psi\vec\tau\gamma^\mu\psi).
 \eeqn
 In principle, one could construct many more higher order interaction terms,
 or derivative terms of higher order, but in practice only a relatively
 small set of free parameters can be adjusted from the data of ground-state
 nuclear properties. The electromagnetic interaction between protons is described
 as usual
 \beq
  {\cal L}^{\rm em}
  = -\frac{1}{4}F^{\mu\nu}F_{\mu\nu}-e\bar\psi\gamma^\mu\dfrac{1-\tau_3}{2}\psi A_\mu,
 \eeq
 where $e$ is the charge unit for protons and it vanishes for neutrons.
 The total Lagrangian density (\ref{Lagrangian}) contains eleven coupling constants
 $\alpha_S$, $\alpha_V$, $\alpha_{TV}$, $\alpha_{TS}$, $\beta_S$, $\gamma_S$, $\gamma_V$, $\delta_S$,
 $\delta_V$, $\delta_{TS}$ and $\delta_{TV}$. The subscripts indicate the symmetry of the couplings:
 $S$ stands for scalar, $V$ for vector, and $T$ for isovector, while the symbol refer
 to the additional distinctions: $\alpha$ refers to four-fermion term, $\delta$ to
 derivative couplings, and $\beta$ and $\gamma$ to the third- and fourth-order
 terms, respectively.

 The pseudoscalar $\gamma_5$ and pseudovector $\gamma_5\gamma_\mu$ channels
 do not contribute at the Hartree level due to the parity conservation in nuclei
 and therefore we have neglected it in the Lagrangian density (\ref{Lagrangian}).
 From the experience of RMF with finite-range (RMF-FR) meson exchange,  a fit, which
 includes the isovector-scalar interaction  has not been found to improve the
 description of nuclear ground state observables. This part of the interaction is
 therefore neglected. Consequently, there are nine free
 parameters in RMF-PC model, which is comparable with those in RMF-FR model.

 Using the mean-field approximation and the ``no-sea'' approximation,
 the operators $ \bar\psi(\hat {\cal O}\Gamma)_i\psi$ in Eq.~(\ref{Lagrangian})
 are replaced by their expectation values and become bilinear forms of
 the Dirac spinor $\psi_k$
 for nucleons
 \beq
 \label{vertices}
  \bar\psi(\hat {\cal O}\Gamma)_i \psi\rightarrow
  \langle \Phi\vert\bar\psi(\hat {\cal O}\Gamma)_i \psi\vert\Phi\rangle
   =\sum_{k }v^2_k\bar\psi_k(\hat {\cal
   O}\Gamma)_i \psi_k,
 \eeq
 where $i$ indicates $S, V$, and $TV$. The sum $\sum\limits_{k}$ runs over only
 positive-energy states with the occupation probabilities $v^2_k$.
 Based on these assumptions, one finds the energy density functional for a
 nuclear system:
 \beq
 \label{Energy}
 E_{\rm DF}[\btau, \rho_S, j^\mu_i, A_\mu]
 = \int d^3r~{\mathcal{E}(\bm{r}}),
 \eeq
 where the energy density
 \beq
 \mathcal{E}(\bm{r})= \mathcal{E}^{\rm kin}(\bm{r})
                  +  \mathcal{E}^{\rm int}(\bm{r})
                  +  \mathcal{E}^{\rm em}(\bm{r})
\eeq%
has a kinetic part
 \beq
 \label{kinetic}
 \mathcal{E}^{\rm kin}(\bm{r}) = \tau(\bm{r})=
  \sum_k\,v_k^2~{\psi^\dagger_k (\bm{r})
    \left(\bm{\alpha}\bm{p} + \beta m-m\right )\psi_k(\bm{r})},
\eeq%
an interaction part
 \beqn
 \label{E12}
 \mathcal{E}^{\rm int}(\bm{r})
 &=& \frac{\alpha_S}{2}\rho_S^2+\frac{\beta_S}{3}\rho_S^3
    + \frac{\gamma_S}{4}\rho_S^4+\frac{\delta_S}{2}\rho_S\triangle \rho_S\nonumber \\
 &&+ \frac{\alpha_V}{2}j_\mu j^\mu + \frac{\gamma_V}{4}(j_\mu j^\mu)^2 +
       \frac{\delta_V}{2}j_\mu\triangle j^\mu  \\
 && +  \frac{\alpha_{TV}}{2}\vec j^{\mu}_{TV}\cdot(\vec j_{TV})_\mu+\frac{\delta_{TV}}{2}
    \vec j^\mu_{TV}\cdot\triangle(\vec j_{TV})_{\mu},\nonumber
\eeqn%
 which contains the local densities and currents
\bsub%
\label{currents}
\beqn%
  \label{E13a}
  \rho_S(\bm{r})          &=&\sum_{k }v^2_k\bar\psi_k(\bm{r})\psi_k(\bm{r}),\\
  \label{E13b}
  j^\mu_{V}(\bm{r})       &=&\sum_{k }v^2_k\bar\psi_k(\bm{r})\gamma^\mu\psi_k(\bm{r}),\\
  \label{E13c}
  \vec j^\mu_{TV}(\bm{r}) &=&\sum_{k }v^2_k\bar\psi_k(\bm{r})\vec\tau\gamma^\mu\psi_k(\bm{r}).
\eeqn%
\esub%
 and an electromagnetic part
 \beq
 \mathcal{E}^{\rm em}(\bm{r})
 =%
 \frac{1}{4}F_{\mu\nu}F^{\mu\nu}-F^{0\mu}\partial_0A_\mu
 +eA_\mu j^\mu_p.
\eeq%

 Minimization of the energy density functional (\ref{Energy}) with respect to $\bar\psi_k$
 gives rise to the Dirac equation (i.e., Kohn-Sham equation) for the single nucleons
 \beqn
  \label{DiracEq}
  [\gamma_\mu(i\partial^\mu-V^\mu)-(m+S)]\psi_k=0.
 \eeqn
 The single-particle effective Hamiltonian contains local scalar $S(\bm{r})$ and vector
 $V^\mu(\bm{r})$ potentials
 \beq
 \label{potential}
   S(\bm{r})    =\Sigma_S, \quad
   V^\mu(\bm{r})=\Sigma^\mu+\vec\tau\cdot\vec\Sigma^\mu_{TV},
 \eeq
 where the nucleon scalar-isoscalar $\Sigma_S$, vector-isoscalar $\Sigma^\mu$ and
 vector-isovector $\vec\Sigma^\mu_{TV}$ self-energies are given in terms of the various
 densities
 \bsub\beqn
  \Sigma_S           &=&\alpha_S\rho_S+\beta_S\rho^2_S+\gamma_S\rho^3_S+\delta_S\triangle\rho_S,\\
  \Sigma^\mu         &=&\alpha_Vj^\mu_V +\gamma_V (j^\mu_V)^3
                       +\delta_V\triangle j^\mu_V + e A^\mu,\\
  \vec\Sigma^\mu_{TV}&=& \alpha_{TV}\vec j^\mu_{TV}+\delta_{TV}\triangle\vec j^\mu_{TV}.
 \eeqn\esub
 For ground state of an even-even nucleus one has time reversal symmetry and
 the space-like components of the currents $\bj_i$ in Eq.~(\ref{currents}) and
 the spatial part of the vector potential $\bV(\bm{r})$ in
 Eq.~(\ref{potential}) vanish. Moreover, because of charge conservation in nuclei,
 only the 3rd-component of isovector potentials $\vec\Sigma^\mu_{TV}$ contributes.
 The Coulomb field $A_0$ is determined by Poisson's equation.

 In addition to the self-consistent mean-field potentials, for open-shell nuclei,
 pairing correlations are taken into account by the BCS method with a smooth cutoff
 factor $f_k$ to simulate the effects of
 finite-range~\cite{Krieger90,Bender00p}, i.e. we have to add to the
 functional (\ref{Energy}) a pairing energy depending on the pairing tensor $\kappa$
 of the form
 \beq
 \label{PairingE}
 E_{\rm pair}[\kappa,\kappa^*]
 = \sum_{kk'>0} f_{k^{}} f_{k'}\langle k {\bar k} \vert V^{pp} \vert k'{\bar k}'\rangle
 \kappa^\ast_{k} \kappa_{k'}.
 \eeq
with the smooth cut-off weight factors %
\beq%
\label{Weight} f_k  =\frac{1}{1+\exp[( \epsilon_k- \epsilon_F-\Delta
E_\tau)/\mu_\tau]},
\eeq%
 where $\epsilon_k$ is the eigenvalue of the self-consistent single-particle field.
 $\epsilon_F$ is the chemical potential determined through the constraint on
 average particle number: $\langle \Phi\vert \hat N_\tau \vert \Phi \rangle = N_\tau$.
 The cut-off parameters $\Delta E_\tau$
 and $\mu_\tau=\Delta E_\tau/10$ are chosen in such a way that
 $2\displaystyle\sum_{k>0}f_k = N_\tau +1.65N^{2/3}_\tau$, where
 $N_\tau$ is the particle number of neutron or proton.

 In the following calculations we use both a mono\-pole force and
 a density-independent $\delta$-force in the pairing channel
 respectively. In the case of the monopole
 force we have $\kappa_k=u_kv_k$ and
 \beq
 \label{monopole}
 E_{\rm pair}[\kappa,\kappa^*]
  = - \sum_{\tau=n,p} G_\tau \left|\sum_{k>0} f_{k} u_{k} v_{k}\right|^2.
\eeq%
 In the case of a $\delta$-force we use
 \beq
 \label{PairingDFT}
 E_{\rm pair}[\kappa,\kappa^*]
 = - \sum_{\tau=n,p} \dfrac{V_\tau}{4}\int d^3r\kappa^\ast_\tau(\bm{r}) \kappa_\tau(\bm{r}).
 \eeq
 where $V_\tau$ is the constant pairing strength and the
 pairing tensor $\kappa(\bm{r})$ is given by
 \beq
  \kappa(\bm{r})
  =-2\sum_{k>0}f_ku_kv_k\vert\psi_k(\bm{r})\vert^2.
 \eeq
The pairing strength parameters $G_{\tau}$ in the case of monopole
pairing and $V_{\tau}$ for zero range pairing forces are adjusted by
fitting the average single-particle pairing gap
\beq%
\langle\Delta\rangle\equiv%
\frac{\sum_{k}f_{k}v_{k}^{2}\Delta_{k}}{\sum_{k}f_{k}v_{k}^{2}}
\label{avgap}%
\eeq%
to the experimental odd-even mass difference obtained with a
five-point formula.

 Moreover, the proper treatment of center of mass (c.m.) motion
 has been found very important in the binding energy of light
 nuclei~\cite{Bender00c,Long04,Chen07}. We adopt the same c.m.
 correction to the total energy after variation, as it has
 been used in adjusting the parameter set PC-F1~\cite{Bender00c},
 \beq
  \label{Eq:Ecm}
  E^{\rm mic}_{\rm cm}=-\dfra{1}{2mA}\langle\hat \bP^{2}_{\rm cm}\rangle,
 \eeq
 where $m$ is the mass of neutron or proton. $A$ is mass number
 and $\hat \bP_{\rm cm}=\sum_i^A \hat \bp_i$ is the total momentum in the
 c.m. frame.

 The total energy for the nuclear system becomes
 \beqn
 E_{\rm tot.}
  = E_{\rm DF}[\btau, \rho_S, j^\mu_i, A_\mu]
   +E_{\rm pair}[\kappa,\kappa^*] +E^{\rm mic}_{\rm cm}.
 \eeqn

 To obtain the potential energy surface (PES), the mass quadrupole moment
 is constrained through the quantities $q_{20}$ and $q_{22}$, which are related
 to the triaxial deformation parameters $\beta$ and $\gamma$ of the Bohr Hamiltonian
 by
 \bsub
 \beqn
 q_{20}&=& \sqrt{\frac{5}{16\pi}}\langle 2z^2-x^2-y^2\rangle = \frac{3}{4\pi}AR^2_0\beta\cos
 \gamma,~~~~~~~~~~~\\
 q_{22}&=& \sqrt{\frac{15}{32\pi}}\langle x^2-y^2\rangle = \frac{3}{4\pi}AR^2_0 \frac{1}{\sqrt{2}}\beta\sin\gamma,
 \eeqn
\esub%
 where $R_0=1.2A^{1/3}$ fm. The total mass quadrupole moment $q$ is
 thus given by
\beq
\label{Q-moment}%
q= \sqrt{\frac{16\pi}{5}} \sqrt{q^2_{20}+2q^2_{22}}.
 \eeq
 We thus obtain mean field wave functions $\vert\Phi(\beta,\gamma)\rangle$ that depend on
 the deformation parameters $\beta$ and $\gamma$. In the following we
 abbreviate the pair of deformation parameters by a single letter
 $q=(\beta,\gamma)$.

 \subsection{Three dimensional angular momentum projection}

 The nuclear mean-field wave function $\vert\Phi\rangle$ is a product of the solutions
 of the deformed Dirac equation of Eq.~(\ref{DiracEq}) and therefore it
 does not have good angular momentum. To obtain the collective energy spectrum and wave
 functions with the good angular momentum $J$, it is crucial to restore the
 spontaneously broken rotational symmetry. Especially, for triaxially deformed
 states $\vert \Phi(q)\rangle$ with the deformation parameters
 $q=(\beta,\gamma)$, a full 3DAMP is
 required.

 The wave function $\vert \Psi^{JM}_{\alpha,q}\rangle$ in the
 laboratory frame, that is an eigenfunction of $\hat J^2$ and $\hat J_z$ with
 the eigenvalues $J(J+1)$ and $M$, is obtained by
projection~\cite{Ring80}
 \beqn
  \label{PRJWF}
  \vert \Psi^{JM}_{\alpha,q}\rangle
  = \sum_K f^{JK}_{\alpha}(q)\vert JMK,q\rangle,
 \eeqn
 where $\alpha=1,2,\cdots$ labels the different collective excited states.
 The basis $\vert JMK,q\rangle$ functions are not just simply Wigner $D$-functions
 as adopted in the classical triaxial rotor model but they are determined
 microscopically from the intrinsic state $\vert\Phi(q)\rangle$ by projection using
 the operators $\hat P^J_{MK}$
 \beq
  \vert JMK,q\rangle=\hat P^J_{MK}\vert\Phi(q)\rangle.
 \eeq
 The projector-like operator $\hat P^J_{MK}$ has the form,
  \beqn
  \hat P^J_{MK}
  =\frac{2J+1}{8\pi^2}\int d\Omega D^{J\ast}_{MK}(\Omega) \hat R(\Omega),
  \eeqn
 with $\Omega$ representing a set of the three Euler angles ($\phi, \theta, \psi$) and the
 measure $d\Omega=d\phi \sin\theta d\theta d\psi$. $D^J_{MK}(\Omega)$ is the Wigner $D$-function
 with the rotational operator chosen in the notation of Edmonds~\cite{Edmonds57} as
 $ \hat R(\Omega)=e^{i\phi\hat J_z}e^{i\theta\hat J_y}e^{i\psi\hat J_z}$.
 The effect of $\hat P^J_{MK}$ is extracting from the intrinsic state $\vert\Phi(q)\rangle$
 the component with an eigenvalue $K$ of the angular momentum projection along the
 intrinsic $z$-axis~\cite{Corbett71,Ring80}. Since $K$ is not a good quantum number for a triaxial shape,
 all these components must be mixed, which corresponds to the so-called ``K-mixing".
 Considering the $D_2$ symmetry of triaxial shape for even-even nuclei, the sum in
 Eq.~(\ref{PRJWF}) is restricted to non-negative even values of $K$. The wave function
 $\vert \Psi^{JM}_{\alpha,q}\rangle$ is therefore simplified as~\cite{Ring80,Burzynski95}
 \beqn
 \label{TrialWF}
  \vert \Psi^{JM}_{\alpha,q}\rangle
  = \sum_{K\geq0}\frac{f^{JK}_{\alpha}(q)}{1+\delta_{K0}}\vert JMK+,q\rangle,%
 \eeqn
 where the angular momentum projected $K$-component, $\vert JMK+,q\rangle$, is given by
 \beq
  \vert JMK+,q\rangle
    =[\hat P^J_{MK}+(-1)^J\hat P^J_{M-K}]\vert\Phi(q)\rangle.
 \eeq
 The expansion coefficients $f^{JK}_{\alpha}(q)$ are determined requiring
 that the energy evaluated on
 $\vert \Psi^{JM}_{\alpha,q}\rangle$ is stationary with respect to
 $f^{JK\ast}_{\alpha}(q)$. This condition leads to the generalized eigenvalue equation
 \beqn
 \label{AMPWHE}
 \sum_{K^\prime\geq0}\{{\cal H}^J_{KK^\prime}(q;q)
 - E^J_\alpha{\cal N}^J_{KK^\prime}(q;q)\}f^{JK^\prime}_{\alpha}(q)=0,
 \eeqn
 where the overlap kernels ${\cal O}^J_{KK^\prime}(q;q)$ are determined
 by ($ {\cal O}={\cal N}, {\cal H}$):
 \beqn
 \label{OverlapK}
  {\cal O}^J_{KK^\prime}(q;q)
 &=&\Delta_{KK^\prime}
     [O^J_{KK^\prime}(q;q)
      +(-1)^{2J}O^J_{-K-K^\prime}(q;q)\nonumber\\
  &&  +(-1)^JO^J_{K-K^\prime}(q;q)
      +(-1)^JO^J_{-KK^\prime}(q;q)],\nonumber\\
 \eeqn
 with $\hat O=1, \hat H$, and $\Delta_{KK^\prime}=1/[(1+\delta_{K0})(1+\delta_{K^\prime0})]$
 \beqn
 \label{Integration1}
  O^J_{KK^\prime}(q;q)
 =\frac{2J+1}{8\pi^2} \int d\Omega D^{J\ast}_{KK^\prime}
     \langle\Phi(q)\vert \hat O\hat R\vert\Phi(q)\rangle.
 \eeqn
 The details about the calculation of overlap functions
 $\langle\Phi(q)\vert\hat O\hat R(\Omega)\vert\Phi(q)\rangle$
 will be given in the next section.

 The generalized eigenvalue equation (\ref{AMPWHE}) is solved in the standard way
 as discussed in Ref.~\cite{Ring80}. It is accomplished by diagonalizing the norm
 kernel ${\cal N}^J_{KK^\prime}(q;q)$ first
 \beq
  \sum_{K^\prime\ge0}{\cal N}^J_{KK^\prime}(q;q)u^{JK^\prime}_m(q)
  =n^{J}_mu^{JK}_m(q).
 \eeq
 The eigenfunctions $u^{JK}_m(q)$ form a complete orthonormalized set
 \bsub
  \beqn
   \sum_mu^{\ast JK}_m(q)  u^{JK^\prime}_m(q)          = \delta_{KK^\prime},\\
   \sum_{K\geq0} u^{\ast JK}_m(q)u^{JK}_{m^\prime}(q)  = \delta_{mm^\prime}.
  \eeqn
 \esub
 The non-zero eigenvalues ($n^J_m\neq0$) of the matrix ${\cal N}^J_{KK^\prime}(q;q)$
 are used to build the normalized vectors (i.e. the natural states) as
 \beq
  \vert m\rangle
 =\frac{1}{\sqrt{n^J_m}}\sum^{J}_{K\geq0} u^{JK}_m(q)\vert JMK+,q\rangle,
 \eeq
 which are orthogonal and define the ``collective" subspace.

 In practice, a cut-off $\chi$ is usually introduced to define the non-zero
 eigenvalues, i.e., $n^J_m>\chi$. In this work, however,
 we do not need such a cut-off. This is because the
 states with zero eigenvalue in norm matrix have already been excluded by constructing
 the collective wave function with the help
 of $D_2$ symmetry as shown in Eq.(\ref{TrialWF}). Of course, if one performs GCM
 calculations, one cannot avoid introducing this cut-off.

 The solution of Eq.~(\ref{AMPWHE}) becomes
 an usual eigenvalue problem,
 \beq
 \label{CollectiveEQ}
  \sum_{m^\prime}\langle m\vert \hat H\vert m^\prime\rangle g^{J\alpha}_{m^\prime}
  = E^J_\alpha g^{J\alpha}_m,
 \eeq
 with the collective Hamiltonian given by the matrix elements
 \beq
 \langle m\vert\hat H\vert m^\prime\rangle
 =\frac{1}{\sqrt{n^J_mn^J_{m^\prime}}}
  \sum_{K,K^\prime\geq0} u^{\ast JK}_m(q)
  {\cal H}^J_{K,K^\prime}u^{JK^\prime}_{m^\prime}(q).
 \eeq
 The solution of Eq.~(\ref{CollectiveEQ}) determines both the energies $E^J_\alpha$
 and the weights $f^{JK}_{\alpha}(q)$ of nuclear states
 $\vert \Psi^{JM}_{\alpha,q}\rangle$,
 \beq
  \displaystyle
  f^{JK}_{\alpha}(q)
   = \sum\limits_{m,n^J_m\neq0}\frac{g^{J\alpha}_m}{\sqrt{n^J_m}}u^{JK}_m(q).
 \eeq

  \subsection{Evaluation of electromagnetic transition probability }

 Once the weights $f^{JK}_{\alpha}(q)$ of nuclear collective wave function
 $\vert \Psi^{JM}_{\alpha,q}\rangle$ are known, it is straightforward to calculate
 all physical observables, such as electromagnetic transition probability.
 Some of them provide a good test of the accuracy of symmetry restoration
 which can be used to determine a sufficient number of mesh points in the
 integration over the Euler angles in Eq.~(\ref{Integration1}). Moreover,
 through the construction of the collective wave function in Eq.~(\ref{TrialWF})
 zero eigenvalues of the norm kernel have been removed.
 There are subsequently $J/2+1$ or $(J-1)/2$
 collective states and rotation energy levels for the even or odd spin $J$~\cite{Enami99}. These
 levels will be assigned into bands according to their B(E2) transition probabilities.

 The $B(E2)$ transition probability from an initial state
 $(q,J_i,\alpha_i)$ to a final state $(q,J_f,\alpha_f)$ is defined by
 \beqn
  B(E2; q,J_i,\alpha_i\rightarrow q,J_f,\alpha_f)
    &=& \frac{e^2}{2J_i+1}
        \vert\langle J_f,q\vert\vert \hat Q_{2}\vert\vert J_i,q\rangle \vert^2.\nonumber\\
 \label{E42}
 \eeqn
 The reduced matrix element of $\langle J_f,q\vert\vert \hat Q_{2}\vert\vert J_i,q\rangle$
 is given by,
 \beqn
 \label{Integration2}
 \langle J_f,q\vert\vert \hat Q_{2}\vert\vert J_i,q\rangle
 =\frac{{\hat J}_i{\hat J_f}}{8\pi^2}
 \sum_{K_iK_f}(-1)^{J_f-K_f}f^{\ast J_fK_f}_{\alpha_f}f^{J_iK_i}_{\alpha_i}\nonumber\\
 \times \sum_{\mu M}
 \left(\begin{array}{ccc}
   J_f  &  2         &J_i \\
   -K_f & \mu        &M \\
 \end{array}
 \right)
 \int d\Omega\, D^{J_i\ast}_{M K_i}
  \langle\Phi(q)\vert
  \hat Q_{2\mu} \hat R\vert\Phi(q)\rangle,\nonumber\\
 \eeqn
with ${\hat J}=2J+1$ and $\hat Q_{2\mu}=r^2Y_{2\mu}$. One can
evaluate the integration over the Euler angles in the interval
$[0,\pi]$ and multiply with the factor
 \beq
 \left[1 +(-1)^{\mu}e^{-iM\pi}
             +e^{-iK_i\pi}
             +(-1)^{\mu} e^{-i(M+K_i)\pi}\right].
  \eeq

 The angular-momentum projection performs a transformation to the
 laboratory frame of reference. This transformation cannot be inverted
 and therefore, an intrinsic deformation
 cannot be unambiguously assigned to the projected states.
 Instead, the comparison between theoretical and experimental ``deformations"
 should be done directly on the basis of B(E2) values and spectroscopic
 quadrupole moments $Q^{(s)}(J,\alpha)$,
 \beqn
 Q^{(s)}(J,\alpha)
 &\equiv&e\sqrt{\dfrac{16\pi}{5}}
 \langle \Psi^{JM=J}_{\alpha,q} \vert \hat Q_{20}\vert \Psi^{JM=J}_{\alpha,q}\rangle\nonumber\\
 &=&e\sqrt{\dfrac{16\pi}{5}}
 \begin{pmatrix}
 J & 2 & J \\
 J & 0 & -J
 \end{pmatrix} \langle J,q \vert\vert \hat Q_2\vert\vert J,q \rangle.
 \eeqn
 Since the $B(E2)$ values and spectroscopic
 quadrupole moments $Q^{(s)}(J,\alpha)$ are calculated in full configuration space, there is no
 need to introduce effective charges, and hence $e$ denotes the bare value of proton
 charge.
 \subsection{Evaluation of the overlap integrals}
 In the following we evaluate the projected matrix elements
 for general many-body operators  $\hat O$ %
 \beqn
 \label{E45}
  O^J_{KK^\prime}
 &=&\frac{2J+1}{8\pi^2} \int d\Omega D^{J\ast}_{KK^\prime}
     \langle\Phi(q)\vert \hat O\hat R\vert\Phi(q)\rangle\\
 &=&\frac{2J+1}{8\pi^2} \int d\Omega\,D^{J\ast}_{KK^\prime}(\Omega)\,
     \langle 0 \vert \hat O \vert\Omega\rangle\,n(\Omega)
 \nonumber
 \eeqn%
 where, for convenience, we have introduced the following notation
 \beq
 \label{E46}
 \langle 0 \vert    \equiv \langle\Phi(q)\vert ,\quad
 \vert \Omega \rangle \equiv \frac{\hat R(\Omega)\vert \Phi(q)\rangle}
                             {n(\Omega)},
 \eeq
with $\langle 0 \vert \Omega \rangle =1$. The rotational overlap
\beq
 \label{E43}
 n(\Omega) = \langle 0 \vert\hat R(\Omega)\vert 0\rangle
\eeq%
is derived in Eq.~(\ref{Norm}) of Appendix~\ref{AppendixA}.

Using the generalized Wick theorem introduced in
Refs.~\cite{Loewdin55,Onishi66,Balian69} the overlap functions
$\langle 0 \vert \hat O \vert\Omega\rangle$ for arbitrary many-body
operators $\hat O$ can be evaluated in terms of the mixed densities
(\ref{mixed})
 \bsub
 \label{mixed1}
 \beqn
 \label{Mixrho}
  \rho_{kl} (\Omega)
  &\equiv& \langle 0\vert a^\dagger_la_k \vert \Omega\rangle,\\
 \label{Mixkap1}
   \kappa^{10}_{kl} (\Omega)
   &\equiv& \langle0\vert  a_la_k \vert \Omega\rangle,\\
 \label{Mixkap2}
    \kappa^{01}_{kl} (\Omega)
    &\equiv& \langle0\vert  a^\dagger_ka^\dagger_l \vert \Omega\rangle^*.
  \eeqn
 \esub
In this way we obtain for instance for a local single particle
operator $Q(\bm{r})$ the projected matrix element%
\beq
  Q^J_{KK^\prime}
 = \int d^3r Q(\bm{r}) \rho^{J}_{KK^\prime} (\bm{r}),
\eeq
 with the projected density
\beq \rho^{J}_{KK^\prime}(\bm{r}) = \frac{2J+1}{8\pi^2} \int d\Omega
D^{J\ast}_{KK^\prime}(\Omega) \rho (\bm{r};\Omega)n(\Omega),
 \eeq
where $\rho (\bm{r};\Omega)$ is the representation of the mixed
density
 (\ref{Mixrho}) in $r$-space given in Eq.~(\ref{mixedr})

For the Hamiltonian overlap in Eq.~(\ref{Integration1}) we find
 \beq
 H^J_{K,K^\prime}=
 \int d^3r~{\cal H}^J_{K,K^\prime}(\bm{r}),%
\eeq%
with%
\beq
 {\cal H}^J_{K,K^\prime}(\bm{r})=
 \frac{2J+1}{8\pi^2} \int d\Omega D^{J\ast}_{KK^\prime}(\Omega)
 {\cal H}(\bm{r};\Omega)n(\Omega),%
\eeq%
where the mixed energy density has the form
\beqn%
{\cal H}(\bm{r};\Omega) &=& {\cal H}^{\rm kin}(\bm{r};\Omega) +
{\cal
H}^{\rm int}(\bm{r};\Omega)\nonumber\\
&&+{\cal H}^{\rm C}(\bm{r};\Omega)+{\cal H}^{\rm
pair}(\bm{r};\Omega).
\eeqn%
The kinetic part
\beq%
\label{Hkin}%
{\cal H}^{\rm kin}(\bm{r},\Omega)= \tau(\bm{r};\Omega)
\eeq%
is given in Eq.~(\ref{B25}). The interaction part ${\cal H}^{\rm
int}(\bm{r},\Omega)$ has the same structure as the corresponding
energy density ${\cal H}^{\rm int}(\bm{r})$ in Eq.~(\ref{E12}). We
only have to replace the densities $\rho(\bm{r})$ and currents
$j^\mu(\bm{r})$ by the mixed densities $\rho(\bm{r};\Omega)$ and the
mixed currents $j^\mu(\bm{r};\Omega)$ derived in Eqs.~(\ref{mixedr})
and (\ref{mixedj}). This is an ad-hoc procedure that is used by
analogy to the Hamiltonian case~\cite{LDB.08}.

The Coulomb part of the mixed energy density has the form
\beq%
{\cal H}^{\rm C}(\bm{r};\Omega)=
\frac{e^2}{8\pi}\rho_p(\bm{r};\Omega)\int d^3r^\prime\,\frac{
\rho_p(\bm{r}^\prime;\Omega)}{|\bm{r}-\bm{r}^\prime|},
\eeq%
 Since the exchange term of Coulomb interaction
 has not been included in the parameterizations of
 relativistic mean-field energy density functional,
 it has been neglected in the energy kernel as well.

Because of time reversal invariance the spatial parts of the
currents $\bj_{V}(\bm{r})$ in Eq.~(\ref{E13b}), $\bj_{TV}(\bm{r})$
in Eq.~(\ref{E13c}) and the electromagnetic current $\bj_{\rm
em}(\bm{r})$ vanish in the mean field calculations. This is no
longer true for the mixed currents in Eq. (\ref{mixedj}). Because of
time reversal symmetry they are purely imaginary. In the present
calculations we take into account $\bj_{V}(\bm{r};\Omega)$ and
$\bj_{TV}(\bm{r};\Omega)$ but, for simplicity, we neglect the small
contributions of the gradient terms of the mixed spatial currents
$\Delta\bj_{V}(\bm{r};\Omega)$ and $\Delta\bj_{TV}(\bm{r};\Omega)$
in Eq. (\ref{E12}) and the mixed electromagnetic current $\bj_{\rm
em}(\bm{r};\Omega)$.

The pairing part for the $\delta$-force is given by
\beq%
{\cal H}^{\rm pair}_\tau(\bm{r};\Omega)= -\,\dfrac{V_\tau}{4}\,
\kap^{01\ast}_\tau(\bm{r};\Omega)\kap^{10}_\tau(\bm{r};\Omega),
\eeq%
where the mixed pairing tensors in coordinate space
$\kap^{01\ast}_\tau(\bm{r};\Omega)$ and
$\kap^{10}_\tau(\bm{r};\Omega)$ are given in Eq.~(\ref{mixedk}). For
the mo\-no\-po\-le force we have
 \beq
{\cal H}_\tau^{\rm pair}(\Omega)
  = -\,G_\tau\,\sum_{k>0}  \kap^{01*}_{k\bar k}(\Omega)
               \sum_{k^\prime>0}  \kap^{10}_{k^\prime\bar k^\prime}(\Omega),
\eeq%
where the mixed pairing densities $\kap^{01*}_{k\bar k}(\Omega)$ and
$\kap^{10}_{k^\prime\bar k^\prime}(\Omega)$  in oscillator space are
given in Eq.~(\ref{C13}).

 The c.m. correction in Eq.~(\ref{Eq:Ecm}) is evaluated only within the mean
 field approximation at each value of $q$. The quality of this approximation
 has not been investigated so far. In this case, the contribution from
 the center-of-mass motion to the energy levels of different spin is the
 same at a fixed deformation.

 \subsection{Symmetries of the overlap integrals}

 The imposed symmetries ($D_{2}$ symmetry and time reversal symmetry)
 in the mean-field calculations give rise to symmetries in the overlaps
 $\langle\Phi(q)\vert \hat O \hat R(\Omega) \vert\Phi(q)\rangle $ and
 allow the reduction of the integration intervals for the Euler angles
 approximate by a factor of 16~\cite{Hara82,Enami99}.

 Specifically, the imposed $D_2$ symmetry reduces the integration intervals for
 the Euler angles $(\phi,\theta,\psi)$ in Eqs.~(\ref{Integration1}) and (\ref{Integration2})
 to $\phi\in[0,\pi]$, $\theta\in[0, \pi]$, $\psi\in[0,\pi]$.
 The symmetries associated with the angles $\phi,\psi$ for the Hamiltonian overlap
 are summarized as follows:
  \bsub%
  \beqn%
  \langle \hat H\hat R(\phi,\theta,\psi)\rangle^\ast
  &=&\langle \hat H\hat R(\psi,\theta,\phi)\rangle,\\
  \langle \hat H\hat R(\phi,\theta,\psi)\rangle^\ast
  &=&\langle \hat H\hat R(\pi-\phi,\theta,\pi-\psi)\rangle.
  \eeqn%
  \esub%
 Therefore we have to calculate the Hamiltonian and norm overlaps
 for the Euler angles $\phi, \psi$ explicitly only in two regions:
 a triangle area with $\psi\in[0,\pi/2], \phi\in[0,\psi]$ and
 a square area with $\psi\in[\pi/2,\pi], \phi\in[0,\pi/2]$.
 Using the above mentioned symmetries we obtain the values
 in the remaining regions.

 For the overlaps of an irreducible tensor operator $\hat T_{\lambda\mu}$, one has
 the following relationships:
 \bsub%
 \beqn%
  \langle \hat T_{\lambda\mu}\hat R(\pi+\phi,\theta,\psi)\rangle
  &=&(-1)^\mu\langle \hat T_{\lambda\mu}\hat R(\phi,\theta,\psi)\rangle,~~~~~~~~\\
  \langle \hat T_{\lambda\mu}\hat R(\phi,\theta,\pi+\psi)\rangle
  &=& \langle \hat T_{\lambda\mu}\hat R(\phi,\theta,\psi)\rangle,\\
  \langle \hat T_{\lambda\mu}\hat R(\pi-\phi,\theta,\pi-\psi)\rangle%
  &=&(-1)^\lambda\langle \hat T_{\lambda-\mu}\hat R(\phi,\theta,\psi)\rangle.%
 \eeqn
 \esub%
 The symmetries associated with $\theta$ are summarized as follows:
 \bsub%
 \beqn%
  \langle \hat H\hat R(\phi,\pi-\theta,\psi)\rangle
      &=& \langle \hat H\hat R(\phi,\theta,\psi)\rangle,~~~~~~~~~~~~~~~~~\\
     \langle \hat T_{\lambda\mu}\hat R(\phi,\pi-\theta,\pi-\psi)\rangle
      &=&(-1)^\mu\langle \hat T_{\lambda\mu}\hat R(\phi,\theta,\psi)\rangle.
  \eeqn
  \esub%

  Details on the derivation of symmetry properties of the overlap
  integrals can be found in Refs.~\cite{Hara82,Enami99} and in Appendix~\ref{AppendixC}.

The restoration of broken symmetries in density functional theory is
connected with spurious divergencies, which have been observed in
connection with number projection by the Madrid group in
Ref.~\cite{AER.01} and in connection with the GCM-method in
Ref.~\cite{Doenau98}. Divergencies have also been noticed in the
calculation of overlap matrix elements between zero-quasiparticle
states and two-quasiparticle states in Ref.~\cite{Tajima92}. The
spurious divergencies in number projection are connected with level
crossings and occur in gauge space at the value of the gauge angle
$\varphi=\pi/2$ for levels with the BCS occupation numbers
$v^2_k=\frac{1}{2}$. These poles do not occur in theories based on
one density independent many-body Hamiltonian, if all the terms in
the projected energy are taken into account in a consistent way, in
particular Fock terms, contributions of the Coulomb and spin-orbit
potential to pairing etc (for details see Ref.~\cite{AER.01}). This
is obviously not the case in most versions of density functional
theory, as for instance in Skyrme or Gogny functionals with
fractional density dependence~\cite{BH.08} or for all cases, where
the effective particle-particle interaction is different from the
effective particle-hole interaction. Covariant density functional
theory, as it is used here, is such a case and such poles have been
found in connection with number projection before the variation in
relativistic theories too~\cite{Lop.02}. In principle the many-body
terms of the point coupling Lagrangian in Eq.~(\ref{E5}) lead to
integer powers of the density dependence, but the Fock terms are
neglected and the pairing part of the density functional cannot be
derived from the same Hamiltonian as the mean field part. In fact,
most of the successful density functionals in the literature have
the problem of such poles. They cause in particular problems in the
case of projection before the variation~\cite{AER.01,DSN.07}. In
addition, the prescription for the evaluation of mixed energy
density in analogy with the generalized Wick's theory for
Hamiltonian based case will also lead finite spurious contributions.

During the years several recipes have been developed to deal with
these problems. The most simple method to avoid the spurious
divergencies is by avoiding the pole in the integration over the
angles, i.e. by avoiding the value $\varphi=\pi/2$ in the case of
number projection. Of course, this does not help for a very fine
integration mesh. One therefore has to look for a plateau in the
projected energy as a function of the number of mesh points. More
recently a method has been developed in Ref.~\cite{LDB.08} where the
projected energy functional is modified and the terms containing the
dangerous level crossings and leading to finite spurious
contributions are removed.

In the present investigations we have not observed the spurious
divergencies. In particular we have found convergence in the number
of mesh points (see Figs.~\ref{fig7}, \ref{fig8}, and~\ref{fig9} of
section~\ref{Sec.IIIB}) and therefore the plateau condition is
fulfilled here.  This might be connected to the fact that we do not
carry out a variation after projection. In Ref.~\cite{ZDS.07} such
problems have been observed in the case angular momentum projection
in systems with cranked wave functions and odd particle number. Of
course, it has to be investigated, whether such divergencies can
also occur in systems with time reversal invariance. Work in this
direction is in progress. Moreover, the investigation of correction
from finite spurious contribution is beyond the scope of the present
work and will be postponed in the future study.

 \section{Results and discussion}
 \label{Sec.III}

  \begin{table*}[tp]
   \centering
   \tabcolsep=8pt
   \caption{The binding energies $E_{\rm B}$ (in MeV), charge radii $R_{\rm C}$
   (in fm) calculated by the triaxially (Tri.) deformed and by the spherical (Sph.)
   RMF-PC codes using the parameter set
   PC-F1 in comparison with the available data. Pairing
   correlation is taken into account by the BCS method with $\delta$-forces.
   In the triaxial calculations the oscillator shell number is chosen as $N_{\rm sh} = 12$
   except for $^{208}$Pb with $N_{\rm sh} = 14$.
   In spherical RMF calculations, both $N_{\rm sh} = 12$ and $N_{\rm sh} = 20$
   are chosen for all nuclei.}
   \begin{tabular}{ c|c|cccccccccc}
   \toprule[1pt]
               &    & $^{16}$O   & $^{40}$Ca& $^{48}$Ca & $^{56}$Ni & $^{112}$Sn  & $^{120}$Sn  & $^{124}$Sn&  $^{132}$Sn & $^{208}$Pb\\
   \hline
   $\Delta_n$  &    & $-$ & $-$ & $-$  & $-$ & +  & +  & +  & $-$ & $-$\\
   \hline
   \multirow{2}*{$E_{\rm B}$}
   & Exp.   &  127.619  &  342.052 &  415.991 &  483.992   & 953.531 & 1020.546 & 1049.963   &  1102.851  &  1636.430\\
   & Tri.   &  127.765  &  344.654 &  415.798 &  480.627   & 952.549 & 1021.010 & 1050.542   &  1103.054  &  1637.300 \\
   & Sph.12 &  127.599  &  344.755 &  415.731 &  480.433   & 952.611 & 1021.087 & 1050.726   &  1102.927  &  1637.768  \\
   & Sph.20 &  127.690  &  345.041 &  416.084 &  480.757   & 953.296 & 1021.636 & 1051.041   &  1103.057  &  1637.241 \\
   \hline
   \multirow{2}*{$R_{\rm C}$}
   & Exp.   & 2.693      & 3.478     & 3.479   & -         & 4.593   &4.655     &4.677  &  -          & 5.504  \\
   & Tri.   & 2.766      & 3.480     & 3.490   & 3.741     & 4.590   &4.644     &4.669  &  4.721      & 5.512  \\
   & Sph.12 & 2.762      & 3.478     & 3.491   & 3.741     & 4.589   &4.643     &4.668  &  4.720      & 5.511  \\
   & Sph.20 & 2.763      & 3.478     & 3.491   & 3.742     & 4.589   &4.642     &4.668  &  4.720      & 5.516  \\
  \bottomrule[1pt]
 \end{tabular}
 \label{tab1}
 \end{table*}

In this section, we discuss 3DAMP+RMF-PC calculations in the nuclei
$^{24}$Mg, $^{30}$Mg and $^{32}$Mg. The intrinsic wave functions
that are  used in the 3DAMP calculation have been obtained as
solutions of the self-consistent RMF equations constrained on the
mass  quadrupole moments. During minimization, parity, $D_{2}$
symmetry, and time reversal symmetry are imposed. The densities are
thus symmetric with respect to reflections on the $x=0$, $y=0$ and
$z=0$ planes. The parameter set chosen for the Lagrangian density in
Eq.~(\ref{Lagrangian})  is PC-F1~\cite{Burvenich02}. The solution of
the equation of motion (\ref{DiracEq})  for the nucleons is
accomplished by an expansion of the Dirac spinors in a  set of
three-dimensional harmonic oscillator basis functions in Cartesian
coordinates with $N_{\mathrm{sh}}$ major shells. The basis is chosen
to be isotropic, i.e. the oscillator parameters are chosen as
$b_{x}=b_{y}=b_{z}=b_{0} = \sqrt{ \hbar/m\omega_{0} } $ in order to
keep the basis closed under rotations~\cite{Egido93,Robledo94}. The
oscillator  frequency is given by $\hbar\omega_{0}=41A^{-1/3}$. The
Poisson's equation  for the electromagnetic field is solved using
the standard Green function method~\cite{VB.72}.

\subsection{Illustrative examples of mean-field calculations}


To illustrate our triaxial RMF-PC mean-field calculation,  the total
binding energies and charge radii of some typical  spherical nuclei,
adopted for adjusting the PC-F1 set, are calculated with triaxially
deformed and spherical RMF-PC approaches with PC-F1 set.  The binding
energies and charge radii, together with the  corresponding data
available are given in Table~\ref{tab1}.

It shows that both the binding energies and the charge radii given by
the triaxially deformed and spherical RMF-PC approaches are in good
agreement with the data. The tiny differences in the  binding
energies by these two approaches are due to the  different numerical
algorithm. Here, we have to point out  that the binding energies of
$^{40}$Ca and $^{56}$Ni  with $N=Z$ are relatively poorly reproduced
with a difference of about $2$-$3$ MeV  which cannot be cured simply
by increasing the shell number $N_{\mathrm{sh}}$ and it may be
ascribed to the missing of proton-neutron pairing correlations in the
present calculations.

 \subsection{Convergence check of three-dimensional angular momentum projection}
 \label{Sec.IIIB}
 %

 \begin{figure}[h!]
  \centering
  \includegraphics[width=6cm]{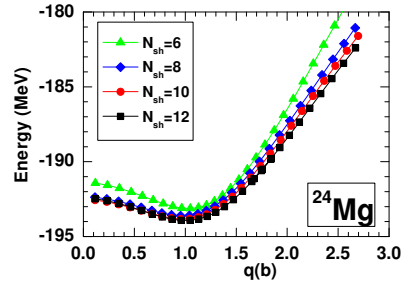}
   \caption{ (Color online) Binding energy curves for $^{24}$Mg,
   calculated by the constrained self-consistent triaxial relativistic mean-field
   approach in a three-dimensional harmonic oscillator basis with major shells $N_{\rm
   sh}=6, 8, 10$, and 12 respectively.}
  \label{fig1}
 \end{figure}

 In Fig.~\ref{fig1}, we show the mean-field binding energy curves for $^{24}$Mg as functions
 of the mass quadrupole moment $q$ ($q_{22}=0$) defined in Eq.~(\ref{Q-moment}),
 calculated by the triaxial RMF-PC approach with the parameter set PC-F1. The four
 different energy curves correspond to the calculations with $N_{\rm sh} =6, 8, 10$, and $12$
 major oscillator shells respectively. It shows that $N_{\rm sh}=8$ is sufficient to obtain
 a reasonably converged mean-field binding energy curve for $^{24}$Mg.
 Pairing correlations have been taken into account by the BCS method
 with mono\-pole pairing forces. The pairing strength parameters $G_\tau$ are determined
 separately for neutrons and protons by adjusting the pairing gaps of the mean-field ground state
 to the odd-even mass difference as obtained with a five-point formula. The pairing
 strength parameters $G_n=34.6/A$ MeV and $G_p=33.75/A$ MeV determined in this
 way have been kept fixed throughout the constraint
 calculations.

 \begin{figure}[h!]
  \centering
  \includegraphics[width=6cm]{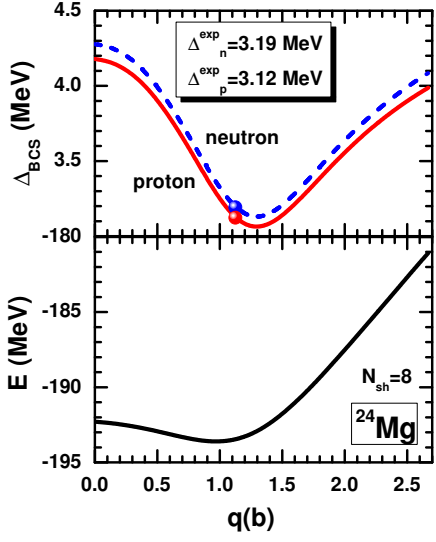}
   \caption{(Color online) Upper panel: Pairing gaps $\Delta_{\tau=n/p}$ of
   neutron (dash line) and proton (solid line). Lower panel:
   energy curve for $^{24}$Mg in a triaxial RMF-PC+BCS calculation with
   a constant pairing strength $G_\tau $, determined by fitting the ground state
   gaps $\Delta_\tau$ to the odd-even mass difference.}
   \label{fig2}
 \end{figure}

In Fig.~\ref{fig2} we plot the pairing gaps of neutrons and protons
in $^{24}$Mg as functions of the quadrupole moment $q$ ($q_{22}=0$)
together with the corresponding energy curve. The total energy shows
a prolate deformed minimum in the energy curve at $q=1.04$ with
$E_{\mathrm{tot.}}=-193.57$~MeV. This figure indicates clearly that
the pairing gap changes considerably with the deformation reflecting
the changes in the single particle level density. Obviously the
minimum in the energy corresponds to a rather low level
density~\cite{Ring80}.

For an axially symmetric intrinsic state, the norm overlap in
Eq.~(\ref{Norm-OVLAP})  can be calculated analytically using the
Gaussian Overlap Approximation  (GOA)~\cite{Onishi66,Beck70}:%
\beq
\label{GOA}%
n(q,q;0,\theta,0) \approx\exp[-\frac{1}{2}\langle\hat J^{2}%
_{y}\rangle\sin^{2}\theta],
\eeq%
which turns out to be an excellent
approximation and thus provides a very  useful test of the numerical
procedure used in angular momentum projection~\cite{Niksic06I}.


 \begin{figure}[tp]
  \centering
  \includegraphics[width=7cm]{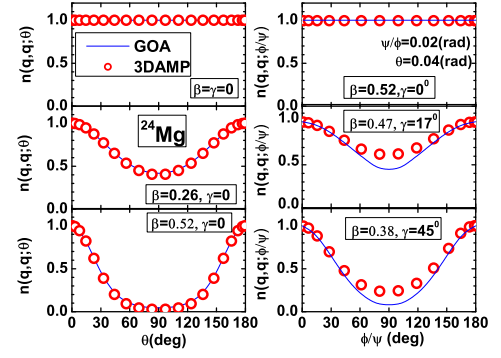}
  \caption{ (Color online) Left panel: A comparison between the norm overlaps
   $n(q,q;\theta)$ as functions of the Euler angle $\theta$ for several
   intrinsic states of $^{24}$Mg obtained by a 3DAMP calculation (open circle)
   and the GOA formula (solid curve).
   Right panel: A comparison between the norm overlaps
   $n(q,q;\phi/\psi)$ as functions of the Euler angles $\phi$ and $\psi$ for  several
   different intrinsic states of $^{24}$Mg obtained by a 3DAMP calculation (open circle)
   and the GOA formula (solid curve), where
   the Eulers angle are $\psi=0.02$ and $\phi=0.02$ radian with $\theta=0.04$ radian.}
   \label{fig3}
 \end{figure}

 Fig.~\ref{fig3} displays the norm overlaps $n(q,q;0,\theta,0)$ as
 functions of the Euler angle $\theta$ for several different axially deformed
 intrinsic states of $^{24}$Mg. It shows that the 3DAMP calculated values
 of the function $n(q;\theta)$ are in good agreement with those given by
 the GOA approximation.

 \begin{figure}[h!]
  \centering
  \includegraphics[width=7cm]{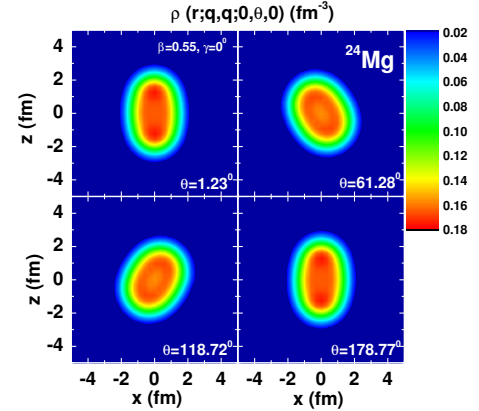}
   \caption{(Color online) The mixed densities $\rho(\bm{r};q,q;\phi,\theta,\psi)$ in $x$-$z$ plane
   with $\beta=0.55, \gamma=0^\circ$.
   The Euler angle $\theta$ has the value of $1.23^\circ$,
   $61.28^\circ$, $118.72^\circ$, and $178.77^\circ$ respectively keeping
   $\phi=\psi=0$.}
   \label{fig4}
 \end{figure}

 \begin{figure}[h!]
  \centering
  \includegraphics[width=7cm]{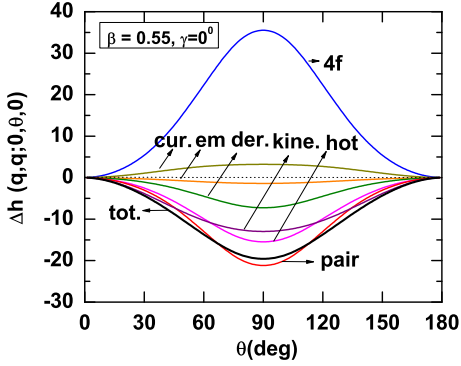}
   \caption{(Color online) The different terms of the Hamiltonian overlap
   for $\beta=0.55$ and $\gamma=0^\circ$ as functions of the Euler angle
   $\theta$ normalized to $\theta=0^\circ$.  }
   \label{fig5}
 \end{figure}

 For triaxially well-deformed intrinsic states, the norm overlap has been
 derived approximately in Refs.~\cite{Beck70,Islam79}:
 \beqn
 \label{GOA_phi}
 n(q,q;\phi,\theta,\psi)
 &\approx&\exp[-\frac{1}{2}\langle
  \hat J^2_y\rangle\theta^2+(\cos(\phi+\psi)-1)\langle \hat J^2_z\rangle\nonumber\\
 && +\frac{i}{2}\langle \hat J_x\rangle\theta(\sin\phi-\sin\psi)].
 \eeqn
In our calculation for $^{24}$Mg, the third term in the exponential
vanishes because of time reversal invariance $\langle \hat
J_x\rangle=0$. The norm overlaps $n(q,q;\phi/\psi)$ are given in
Fig.~\ref{fig3}  as functions of the Euler angles $\phi$ and $\psi$
for several triaxially deformed intrinsic states of $^{24}$Mg. It is
found that the norm overlaps $n(q,q;\phi/\psi)$ oscillate in an exact
3DAMP calculation as functions of $\phi$ and $\psi$ with a
period of $T=180^{\circ}$.  The approximate formula Eq.~(\ref{GOA_phi}%
) is obviously valid only in the  interval $0^{\circ}$ to
$90^{\circ}$. In order to obtain the approximate results in the
interval between $90^{\circ}$ and $180^{\circ}$ we use symmetry
around the angle $\phi/\psi=90^{\circ}$. In this  case,
Fig.~\ref{fig3} shows that the Gaussian overlap approximation can
roughly reproduce the results obtained by the exact 3DAMP
calculations.  Moreover, as expected, the larger the $\gamma$
deformation of the intrinsic state is,  the larger is the amplitude
of the oscillating norm overlaps $n(q;\phi/\psi)$.

 \begin{figure}[h!]
  \centering
  \includegraphics[width=7cm]{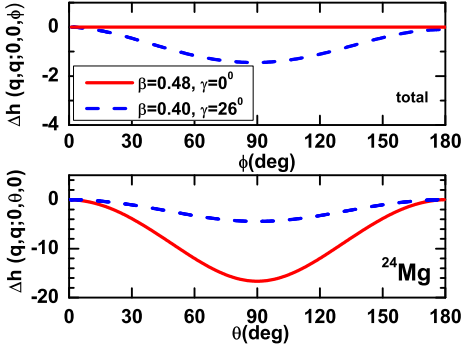}
   \caption{(Color online) Upper panel: Hamiltonian overlap
   with $\beta=0.48, \gamma=0^\circ$ as a function of the Euler angle
   $\theta$.
   Lower panel: Hamiltonian overlap with $\beta=0.40, \gamma=26^\circ$
   as a function of the Euler angles $\phi$ and $\psi$.
   The values with $\theta=0$ or $\phi$( or $\psi)=0$ are chosen as zero.  }
   \label{fig6}
 \end{figure}

 \begin{figure*}[]
  \centering
  \includegraphics[width=12cm]{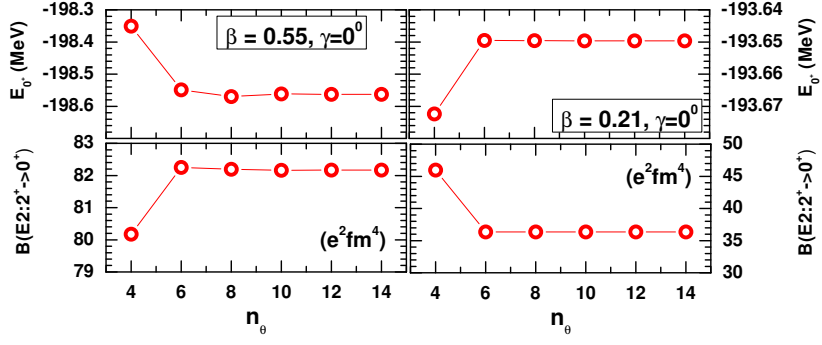}
   \caption{(Color online) The projected energy of the lowest $0^+$ state derived
   from mean-field states with $\beta=0.55, \gamma=0$
   and $\beta=0.21, \gamma=0^\circ$ for the nucleus $^{24}$Mg,
   and the $B(E2\downarrow: 2^+\rightarrow 0^+)$ transition
   probability, as functions of the number of mesh points $n_\theta$
   for the Euler angle $\theta$. }
   \label{fig7}
 \end{figure*}

To describe the collective motion of nuclei in the context of energy
density functional theory, one should determine the corresponding
collective Hamiltonian. In the 3DAMP+RMF-PC approach, the matrix
elements of collective Hamiltonian are constructed in
Eq.~(\ref{OverlapK}) in terms of the Hamiltonian overlaps, which have
their standard functional form but depend upon the mixed densities
and currents.

In Fig.~\ref{fig4}, we plot the mixed nucleon densities $\rho({\bm{r}%
};q,q;\phi,\theta,\psi)$ in the $x$-$z$ plane derived from the
mean-field state with $\beta=0.55,\gamma=0^{\circ}$ for
$\phi=\psi=0^{\circ}$ and for various Euler angles $\theta=$
$1.23^{\circ}$, $61.28^{\circ}$, $118.72^{\circ }$, and
$178.77^{\circ}$. It is obvious that the reflection symmetries with
respect to the planes $x=0$, $y=0$ and $z=0$ present in the
mean-field densities are violated in the corresponding mixed
densities. Moreover, we show in Fig.~\ref{fig5} the various terms in
the Hamiltonian overlap
$h(q_{a},q_{a};\Omega)\equiv{}_{a}\langle0|\hat
{H}|\Omega\rangle_{a}$ resulting from the four-fermion coupling
term, the current contributions, the Coulomb term, the derivative
term, the kinetic term, the higher order term and the pairing term
as functions of the Euler angle $\theta$ for the mean-field state at
the point $\beta=0.55,\gamma =0^{\circ}$. The energy surface is
normalized to $\theta=0$, i.e. $\Delta
h(q_{a},q_{a};\Omega)=h(q_{a},q_{a};\Omega)-h(q_{a},q_{a};0)$. We
find that the current contributions and the Coulomb term in the
Hamiltonian overlap change mildly with the rotation angle $\theta$
and thus they have only small contributions to the collective
Hamiltonian. On the contrary, the four-fermion coupling term, the
pairing term and the higher order term are sensitive to the Euler
angle $\theta$ and play a dominant role in the collective
Hamiltonian.

 In Fig.~\ref{fig6}, we display the total Hamiltonian overlap for the
 axially deformed mean-field state with $\beta=0.48, \gamma=0^\circ$
 and the triaxially deformed mean-field state with $\beta=0.40, \gamma=26^\circ$
 as functions of the Euler angle $\theta$, or the Euler angles $\phi$ and $\psi$.
 It shows that both for the axially deformed shape
 and the triaxially deformed shape, the Hamiltonian overlaps, behaving like
 the norm overlaps, oscillate with the period $T=180^\circ$ in the Euler angle $\theta$,
 $\phi$, or $\psi$.

A N-point Gaussian-Legendre quadrature is used for integration over
the Euler angles  $\phi,\theta$ and $\psi$ in the calculations of the
norm kernel $\mathcal{N}^{J}_{KK^{\prime}}$  and the Hamiltonian
kernel $\mathcal{H}^{J}_{KK^{\prime}}$. The calculation of the
Hamiltonian overlap  at each mesh point of the Euler angles is very
time consuming. Therefore, besides the utilization  of symmetries in
overlaps, it is essential to make a careful check of the convergence
for the number of mesh points. The projected energy and the $B(E2)$
transition probability  are good observables for this purpose.

 \begin{figure}[h!]
  \centering
  \includegraphics[width=9cm]{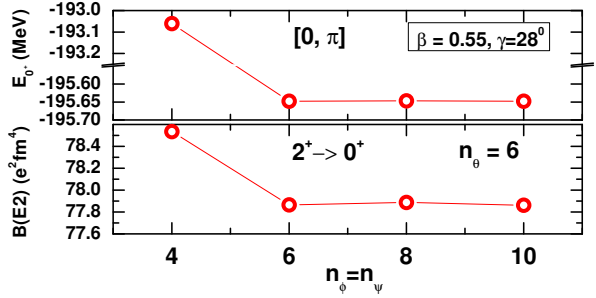}
   \caption{(Color online)  The projected energy of the first $0^+$ state obtained
   from the mean-field state with $\beta=0.55, \gamma=28^\circ$
   for $^{24}$Mg, and the $B(E2\downarrow: 2^+\rightarrow 0^+)$ transition
   probability, as functions of the number of mesh points $n_\phi$
   (or $n_\psi$) for the Euler angles $\phi$ (or $\psi$). }
   \label{fig8}
 \end{figure}

 In Fig.~\ref{fig7}, we plot the projected energy of first $0^+$ state obtained
 from the mean-field states with $\beta=0.55, \gamma=0^\circ$ and
 $\beta=0.21,\gamma=0^\circ$ for $^{24}$Mg,
 and the corresponding $B(E2\downarrow: 2^+\rightarrow 0^+)$ transition
 probabilities as functions of the number of mesh points $n_\theta$ for the Euler angle
 $\theta$. The projected energy of the $0^+$ state from the mean-field states with
 $\beta=0.55, \gamma=28^\circ$ and the $B(E2\downarrow: 2^+\rightarrow 0^+)$ transition
 probability, as functions of the number of mesh points $n_\phi$ (or $n_\psi$) for the Euler
 angle $\phi$ (or $\psi$) are shown in Fig.~\ref{fig8}, where $\theta$, $\phi$ and $\psi$
 have values between 0 and $\pi$. We find that in order to achieve a
 precision of $0.001\%$ for $E_{0^+}$ and $0.1\%$ for $B(E2: 2^+\rightarrow 0^+)$
 the total number of mesh points for
 the Euler angles in the intervals $\phi\in[0,\pi]$, $\theta\in[0,\pi]$, $\psi\in[0,\pi]$ should
 fulfil the relation:
 $N_\phi\times N_\theta\times N_\psi\geq6\times6\times6$.

 \begin{figure}[h!]
  \centering
  \includegraphics[width=8cm]{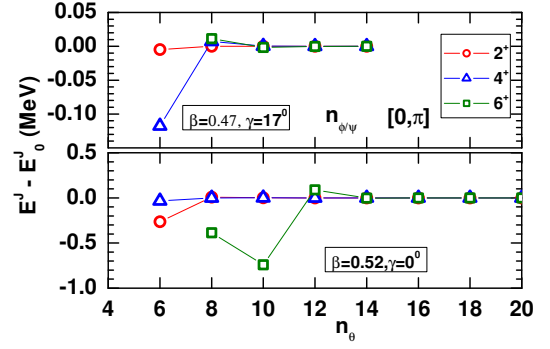}
   \caption{(Color online)  The projected energies of $2^+, 4^+$ and $6^+$ states from
   the mean-field states with $\beta=0.47, \gamma=17^\circ$
   and $\beta=0.52, \gamma=0^\circ$ for $^{24}$Mg as functions of the
   number of mesh points $n_\phi$ (or $n_\psi$) or $n_\theta$.
   $E^J_0$ is the converged energy of a state with spin $J$.}
   \label{fig9}
 \end{figure}

 \begin{figure}[h!]
  \centering
  \includegraphics[width=7cm]{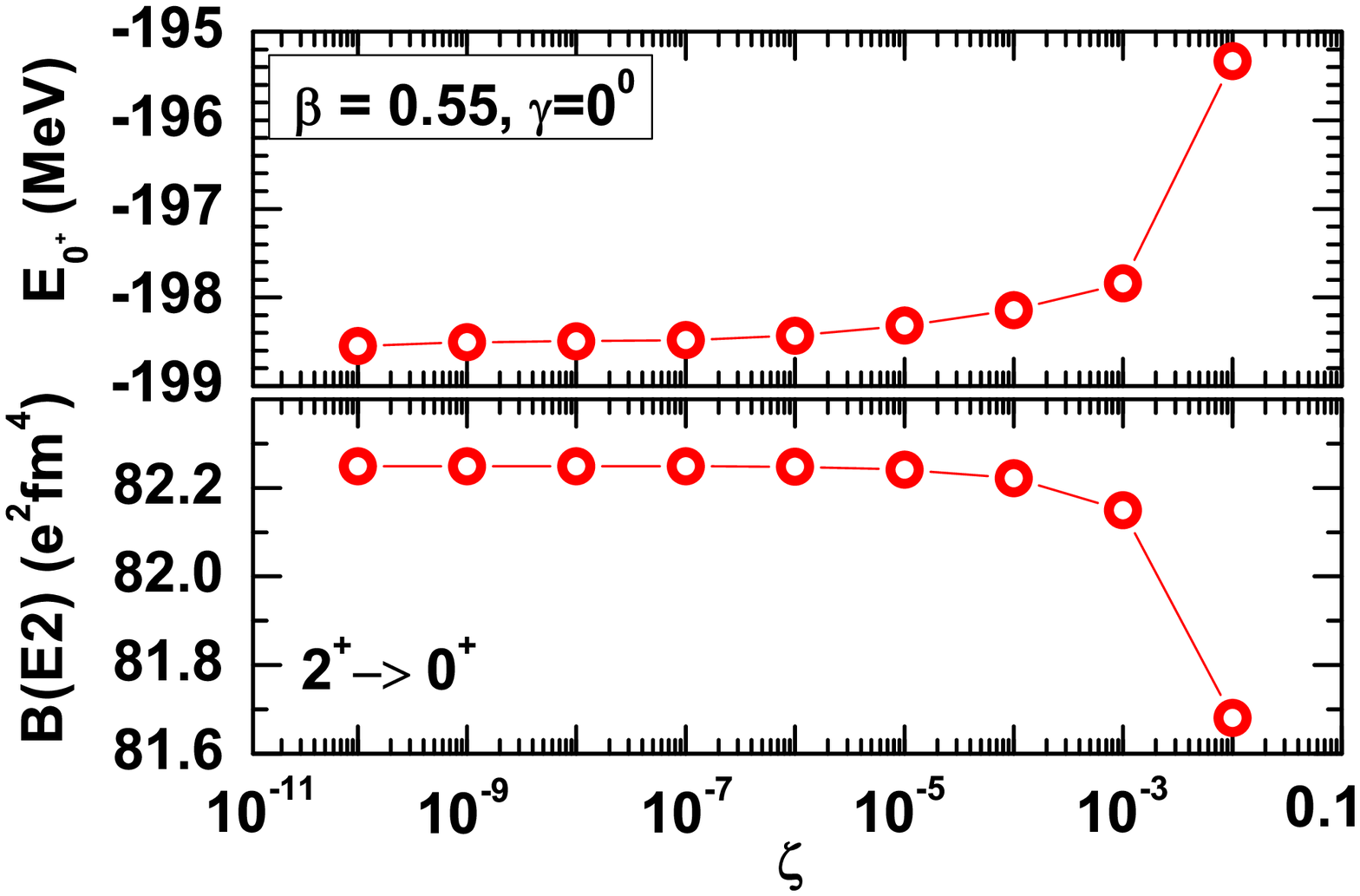}
   \caption{(Color online)  The projected energy of $0^+$ state from
   the mean-field solution with $\beta=0.55, \gamma=0^\circ$ for $^{24}$Mg,
   and the $B(E2\downarrow: 2^+\rightarrow 0^+)$ transition
   probability, as functions of cut-off $\zeta$ in Dirac space ). }
   \label{fig10}
 \end{figure}

 According to the uncertainty principle $\Delta J\cdot\Delta\Omega\simeq\hbar$,
 we need a large number for meshpoints in the Euler angels for higher values of the spin.
 In Fig.~\ref{fig9}, we show the projected energies
 of $2^+, 4^+$ and $6^+$ obtained from the mean-field states with $\beta=0.47, \gamma=17^\circ$
 and $\beta=0.52, \gamma=0^\circ$ as functions of the number of mesh points $n_\phi$,  $n_\psi$
 and $n_\theta$. We find that it is possible with $N_\phi\times N_\theta\times N_\psi\geq12\times14\times12$,
 to achieve a precision of $0.001\%$ in the energy of a projected state
 with angular momentum up to $J=6$ in the ground state band. In the following calculations we
 use such large numbers of mesh points in the Euler angles.

 Since the states with very small occupation probabilities give negligible
 contributions to the kernels, as usual, we introduce a cut-off parameter $\zeta$,
 which divides the Dirac space into an occupied part and an unoccupied part (see
 Eq.~(\ref{Division}). The states
 with $v^2_k\leq\zeta$ will be excluded in the calculation of the overlaps.
 In Fig.~\ref{fig10} we show the projected energy of the first $0^+$ state
 and the $B(E2\downarrow: 2^+\rightarrow 0^+)$ transition probability
 projected from the mean-field state with $\beta=0.55, \gamma=0^\circ$
 as functions of the cut-off parameter $\zeta$. It shows that $\zeta$
 should be chosen as $\zeta\leq10^{-8}$ in order to get a precision of
 $0.01\%$ for $E_{0^+}$ and of $0.00001\%$ for the $B(E2\downarrow:
 2^+\rightarrow 0^+)$ value. Using the cut-off $\zeta$ reduces the computational effort (
 about 80\% of total computer time for $N_{\rm sh}=8$) in the calculations
 of the norm overlap and the matrix elements of mixed densities and
 pairing tensors considerably, especially for the cases of large $N_{\rm sh}$, small particle
 number and weak pairing, where most single particle levels of the Dirac basis have nearly zero
 occupation probabilities.

 \subsection{Tests of three-dimensional angular momentum projection}

 \subsubsection{Application to an axially deformed shape}

 To illustrate the validity of our newly-developed 3DAMP+RMF-PC+BCS code,
 we first apply it to the axially deformed case, where a 1DAMP calculation
 is possible. The projected $J^\pi$ = $0^+$, $2^+$, $4^+$, $6^+$,
 and $8^+$ potential energy curves of $^{32}$Mg have already been calculated
 with 1DAMP+RMF-PC+BCS approach in Ref.~\cite{Niksic06I}. To make a comparison,
 we perform the same calculations within the 3DAMP+RMF-PC+BCS approach. The numerical
 techniques are the same as those of Ref.~\cite{Niksic06I}. We find that
 our newly-developed 3DAMP+RMF-PC+BCS code can reproduce the
 results given by 1DAMP+RMF-PC+BCS approach.

  Furthermore, following Ref.~\cite{Bender08}, we first test the
  3DAMP+RMF-PC approach for an axially deformed shape, which allows two
  distinct orientations in the intrinsic frame: the symmetry axis
  can either parallel to the $z$-axis or it can be perpendicular to it.

 \begin{figure}[h!]
  \centering
  \includegraphics[width=8cm]{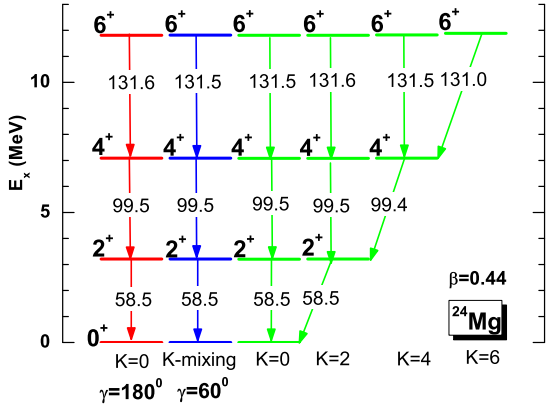}
   \caption{(Color online)  The excitation spectra and $B(E2)$ values projected from
   the axially deformed mean-field states with $\beta=0.44, \gamma=180^0$
   and $\beta=0.44, \gamma=60^0$ respectively.
   The first and second columns show the unique band with $K=0$
   and the unique band with $K$-mixing.
   The last four columns show the decomposition into
   $K$-components when the symmetry axis is chosen perpendicular to the
   $z$ axis, i.e. the $K=0, 2, 4, 6$ bands respectively.}
  \label{fig11}
 \end{figure}

 \begin{figure}[h!]
  \centering
  \includegraphics[width=9cm]{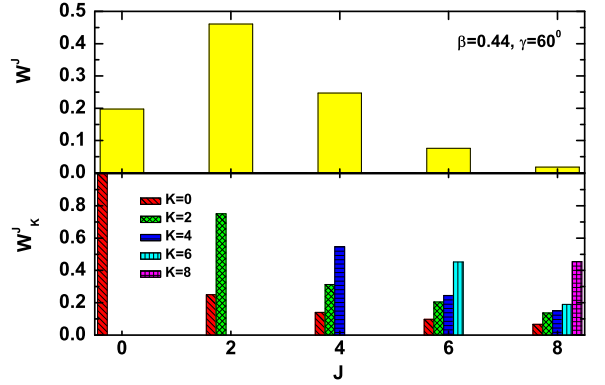}
   \caption{(Color online) Upper panel: The probabilities $W^J$ of finding the component with given spin
   $J$. Lower panel: the probabilities $W^J_K$ of finding the component with
   given spin
   $J$ and the projection $K$ along the $z$-axis in the mean-field state
   with the deformation parameters $\beta=0.44, \gamma=60^\circ$.}
   \label{fig12}
 \end{figure}


 \begin{figure*}[t]
  \centering
  \includegraphics[width=10cm]{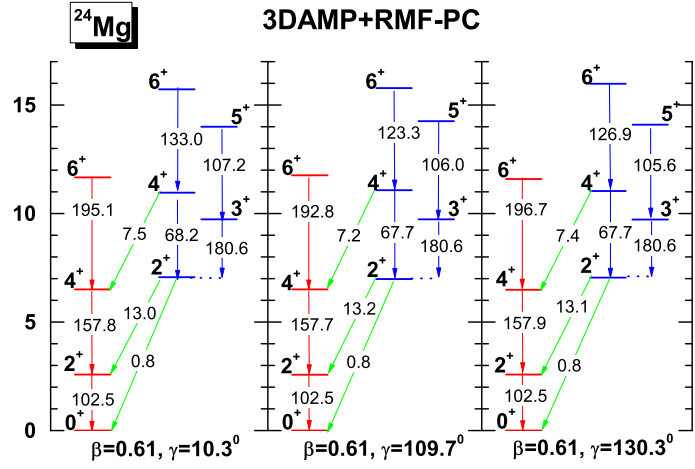}
   \caption{(Color online) The excitation spectra and $B(E2)$ values projected from
   triaxially deformed mean-field states with $\beta=0.61,
   \gamma=10.3^0$; $\beta=0.61, \gamma=109.7^0$;
   and $\beta=0.61, \gamma=130.3^0$ respectively.}
  \label{fig13}
 \end{figure*}

  In Fig.~\ref{fig11}, we show the excitation spectra and $B(E2)$ values
  for $^{24}$Mg projected from the axially deformed mean-field states
  with $\beta=0.44, \gamma=180^0$ and $\beta=0.44, \gamma=60^0$ respectively.
  For $\gamma=180^0$, $z$-axis is along the symmetry axis, and therefore only
  one pure $K=0$ band can be found. All other $K$-components have zero norm.
  While for $\gamma=60^0$, one can show that the pure $K = 0$ state is transformed
  into a multiplet of states with $K$ ranging between 0 and $J$.
  Such phenomena can be seen more clearly from the probabilities
  $W^J\equiv\sum_K\langle \Phi(q)\vert \hat P^J_{KK}\vert\Phi(q)\rangle$
  of finding a component with given spin $J$ and the probabilities
  $W^J_K\equiv\langle \Phi(q)\vert \hat P^J_{KK}\vert\Phi(q)\rangle/W^J$
  of finding a component with given spin $J$ as well as given projection
  $K$ along $z$-axis. These probabilities are shown in Fig.~\ref{fig12}.

  In principle, the transformed wave functions differ only by an unobservable
  phase and the energies of projected states as well as the  electromagnetic
  transition probabilities should be identical. This provides us an excellent
  test of the numerical accuracy of the projection scheme in the code.
  Fig.~\ref{fig11} shows that for the low spin states, e.g., $0^+, 2^+$,
  the projected energies and $B(E2)$ values are exactly the same.
  As angular momentum increases, the difference increases to a
  largest value ($\sim0.4\%$) in the $B(E2\downarrow: 6^+\to4^+)$,
  which could be reduced with more mesh points in the Euler angles.

  \subsubsection{Application to a triaxially deformed shapes}

  The excitation energies and $B(E2)$ values for $^{24}$Mg projected from
  the triaxially deformed mean-field states with $\beta=0.61,\gamma=10.3^0$;
  $\beta=0.61, \gamma=109.7^0$ and $\beta=0.61, \gamma=130.3^0$
  are presented in Fig.~\ref{fig13}. All the excitation energies
  are arranged into bands according to the $B(E2)$ values. These three intrinsic
  states correspond to the same nuclear shape with three different
  orientations in the intrinsic frame. The projected energy and the
  electromagnetic transition probability do not depend on the orientation
  of the nucleus and therefore, in principle, the predicted values should
  be the same as illustrated in Fig.~\ref{fig13}. It shows that the projected energies
  and $B(E2)$ values in these cases are in good agreement with each other.
  However, small differences in the $B(E2)$ values appear and increase with
  angular momentum. Except for the $B(E2: 6^+\rightarrow 4^+)$ in the $K=2$
  band, the difference is smaller than $1\%$. This indicates that more
  mesh points in the Euler angles are necessary to provide a better
  description of the $B(E2: 6^+\rightarrow 4^+)$.

 \begin{figure}[h!]
  \centering
  \includegraphics[width=8cm]{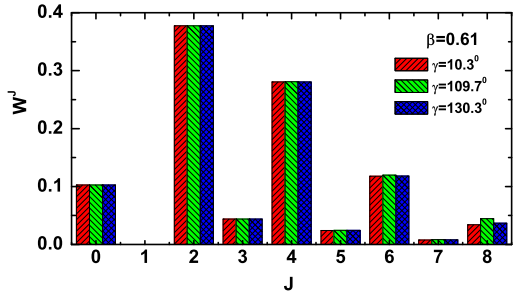}
   \caption{(Color online) The probabilities $W^J$ in the mean-field states with $\beta=0.61,
   \gamma=10.3^0$, $\beta=0.61, \gamma=109.7^0$,
   and $\beta=0.61, \gamma=130.3^0$ respectively.}
   \label{fig14}
 \end{figure}

  The decomposition of a triaxial mean-field state into components with
  different $J$-values in the laboratory frame should also be independent on
  its orientation in the intrinsic frame. In Fig.~\ref{fig14}, we show almost the
  same probabilities $W^J$ of different spin states in these cases.

 \subsubsection{Dispersion of particle numbers}

 \begin{figure}[h!]
  \centering
  \includegraphics[width=8cm]{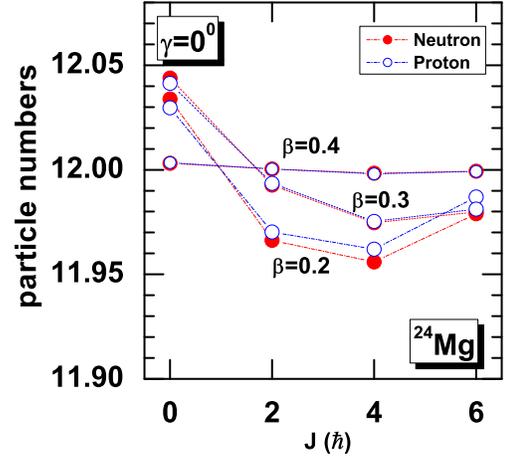}
   \caption{(Color online) The average neutron (filled circle) and proton
    (open circle) numbers of angular momentum projected states with $J\leq6$
   from axially deformed intrinsic states of $^{24}$Mg with $\beta=0.2, 0.3,
   0.4$.}
   \label{fig15}
 \end{figure}
 Although the mean-field intrinsic states are obtained with the constraint
 on the right average particle number, it cannot guarantee the right particle
 number in the angular momentum projected states.
 In order to make up this flaw, in principle, one has to perform PNP calculation.
 The study with both PNP and 3DAMP in the
 context of GCM has only been attempted based on a Skyrme EDF
 theory~\cite{Bender08}. Such kind of study based on a covariant EDF theory
 is still extremely time-consuming. As the first step, in this work, neither the
 exact projection on particle numbers $N$ and $Z$, nor a constrain on the average number of
 particle in the angular-momentum projected states is performed.
 Therefore, it is essential to know the dispersion of particle numbers
 within a rotational band. In Fig.~\ref{fig15}, we plot the average neutron and proton numbers
 of angular momentum projected states with $J\leq6$ from axially deformed
 intrinsic states of $^{24}$Mg with $\beta=0.2, 0.3, 0.4$. It shows that the error in average
 particle number of projected states with $J\leq6$ is within
 $0.5\%$.

 \subsection{Examples of three-dimensional angular momentum projection}

 \subsubsection{Application to $^{24}$Mg}
   \begin{table}[h!]
   \centering
   \tabcolsep=4pt
   \caption{Reduced $E2$ transition probabilities
   from states $J^\pi_i$ to states $J^\pi_f$ in $^{24}$Mg. The minimum of the projected $J=2$ PES
   is used for the calculation of intrinsic wave function.
   The experimental data for the excitation energies $E_{\rm x}$ [in units of MeV]
   and $E2$ transition probabilities [in units of e$^2$fm$^4$] are taken from most recent available
   sources. 1e$^2$fm$^4 = 4\pi(\frac{5}{3})^2(1.2A^{1/3})^{-4}$ W.u.
   $= 0.243$ W.u. for $^{24}$Mg.}
   \begin{tabular}{ c c c c c c }
   \hline\hline
   $J^\pi_i$           &$E_{xi}$(Exp.) &  $J^\pi_f$  &$E_{xf}$(Exp.)    & $B(E2)_{\rm Exp.}$ & $B(E2)_{\rm The.}$ \\
   \hline
   $2^+_1$           & 1.37    &    $0^+_1$  & 0.0        &   $86.4 \pm1.6 $\footnotemark[1]&   74.5 \\
   $4^+_1$           & 4.12    &    $2^+_1$  & 1.37       &   $155.6\pm12.3$\footnotemark[1]&  104.9  \\
   $6^+_1$           & 8.11    &    $4^+_1$  & 4.12       &   $156.4\pm53.5$\footnotemark[1]&  131.3 \\
   \hline
   $2^+_2$           & 4.24    &    $0^+_1$  & 0.0        &   $6.6  \pm0.4 $\footnotemark[1]&  12.3  \\
   $2^+_2$           & 4.24    &    $2^+_1$  & 1.37       &   $12.3 \pm2.1 $\footnotemark[1]&  32.1 \\
   $3^+_2$           & 5.24    &    $2^+_1$  & 1.37       &   $9.5  \pm0.8 $\footnotemark[1]&  21.4 \\
   $3^+_2$           & 5.24    &    $4^+_1$  & 4.12       &   $<17.7$       \footnotemark[2]&  32.5  \\
   $4^+_2$           & 6.01    &    $2^+_1$  & 1.37       &   $4.1 \pm0.8 $\footnotemark[1] &  21.8 \\
   $4^+_2$           & 6.01    &    $4^+_1$  & 4.12       &   $4.1 \pm4.1 $\footnotemark[2] &  21.0 \\
   $6^+_2$           & 9.53    &    $4^+_1$  & 4.12       &   $2.5 \pm1.2 $\footnotemark[1] &  0.4  \\
  \hline
   $3^+_2$           & 5.24    &     $2^+_2$ & 4.24       &   $156.4\pm22.6 $\footnotemark[1]& 134.2 \\
   $4^+_2$           & 6.01    &     $2^+_2$ & 4.24       &   $ 77.0\pm9.9$\footnotemark[1]  & 63.0 \\
   $5^+_2$           & 7.81    &     $3^+_2$ & 5.24       &   $144.0\pm20.3$\footnotemark[1] & 103.7  \\
   $6^+_2$           & 9.53    &     $4^+_2$ & 6.01       &   $ 74.2\pm32.9 $\footnotemark[1]& 44.9 \\
  \hline\hline
 \end{tabular}
 \label{tab2}
 \footnotetext[1]{from Ref.~[\onlinecite{Keinonen89}].}
 \footnotetext[2]{from Ref.~[\onlinecite{Branford75}].}
 \end{table}

The 3DAMP+RMF-PC approach has been used in Ref.~\cite{Yao08cpl2}
 to describe the PES in the $\beta$-$\gamma$ plane for the lowest
 $J^\pi=0^+$ state and for the first excited $J^\pi=2^+$ in the
 nucleus $^{24}$Mg. There is no pronounced minimum with an obvious
 $\gamma$-deformation in the PES for the $0^{+}$ state, which is in
 disagreement with the results of Ref.~\cite{Bender08}, where a 3DAMP+PNP
 calculation  based on a non-relativistic Skyrme HFB functional shows a pronounced
 triaxial minimum with $\beta = 0.6$ and $\gamma=16^\circ$. Keeping in mind that
 we found strong pairing gaps in our mean-field calculations, an
 additional number projection is not expected to change this result.
 A possible reason for this difference is the fact that different
 energy functionals are used in these two calculations.
 A minimum with $\beta\approx0.55,~\gamma\approx10^\circ$ has been found on the PES of
 the first excited $2^{+}$ state. To construct the excitation spectrum and to
 calculate $B(E2)$ transitions, one should in principle perform a GCM configuration
 mixing calculation on top of three-dimensional angular momentum projection, or choose
 the minimum of the different $J$ projected PES as basis. However, such kind of
 calculations are beyond our present study.
 Instead, we use the minimum of the projected $J=2$ PES as basis to calculate the
 experimentally observed excited energy levels and the $B(E2)$ transition probabilities
 in $^{24}$Mg using Eqs.~(\ref{CollectiveEQ}) and (\ref{E42}).
 This is the only way to obtain $K = 2$ bands in our
 calculation. The details about the $B(E2)$ transition probabilities
 in $^{24}$Mg are given in Tab.~\ref{tab2}. It shows that the
 predicted intraband $B(E2)$ values are systematically smaller than
 the data, while the interband $B(E2)$ values are systematically
 overestimated. It indicates that the amplitude of ``K-mixing"
 is too strong in our calculations.

 \begin{figure}[h!]
  \centering
  \includegraphics[width=8cm]{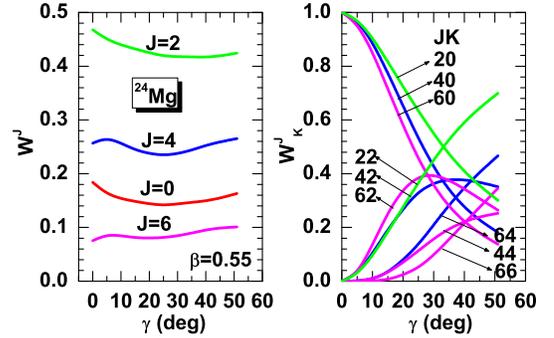}
   \caption{(Color online) The probabilities $W^J$ and the probabilities $W^J_K$ in
   mean-field states with $\beta=0.55$ as functions of triaxial deformation $\gamma$.}
   \label{fig16}
 \end{figure}

 In order to understand the effect of $\gamma$-deformation
 on the amplitude of angular momentum mixing, it is useful to
 investigate the individual components forming the
 intrinsic state $\vert\Phi(q)\rangle$, i.e. the $J$- and
 $K$-mixing. In Fig.~\ref{fig16}, we present the probabilities $W^J$
 and the probabilities $W^J_K$ in the
 mean-field states with $\beta=0.55$ as functions of the triaxiality
 parameter $\gamma$, ranging between $0^\circ$ and
 $60^\circ$. It is noted that each $\gamma$-deformation in this
 range corresponds to a definite shape uniquely.
 Fig.~\ref{fig16} shows that the $J$-mixing remains
 practically constant with changes in the $\gamma$-deformation, while the
 amount of $K$-mixing increases considerably with increasing
 triaxiality. This indicates that the
 underestimated intraband $B(E2)$ values and the overestimated interband $B(E2)$
 in the low-lying excited states of $^{24}$Mg as shown in Tab.~\ref{tab2} are due
 to the large $\gamma$-deformation in the intrinsic state.

\begin{figure}[h]
  \centering
  \includegraphics[width=8cm]{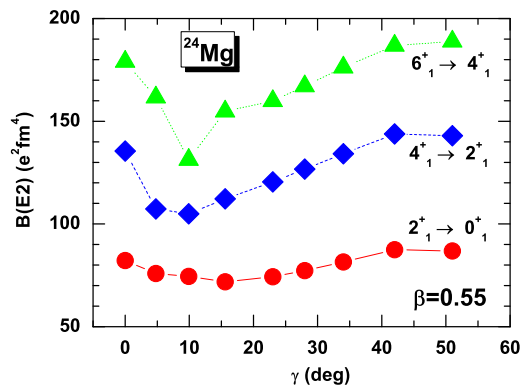}
   \caption{(Color online) $B(E2\downarrow$) transition probabilities for
   $2^+_1 \to 0^+_1$, $4^+_1 \to 2^+_1$
   and $6^+_1 \to 4^+_1$ in the ground state band, projected from
   the mean-field state with $\beta=0.55$ in the nucleus $^{24}$Mg,
   as functions of the triaxial deformation $\gamma$.}
   \label{fig17}
 \end{figure}

 To illustrate the effect of $\gamma$-deformation on the
 $B(E2\downarrow$) values, we plot in Fig.~\ref{fig17} the intraband
 $B(E2\downarrow$) transition probabilities for
 $2^+_1 \to 0^+_1$, $4^+_1 \to 2^+_1$ and $6^+_1 \to 4^+_1$
 in the ground state band projected from mean-field states with $\beta=0.55$
 as functions of the $\gamma$-deformation.
 Obviously the intraband $B(E2\downarrow$) values increase when $\gamma$ approaches
 $0^\circ$ or $60^\circ$. It indicates that a configuration mixing calculation
 (GCM) within a generator coordinate method might be very important to understand
 the observed $B(E2$) values. Alternatively, calculating $B(E2)$ value using the minima of
 each $J$ projected PES might also improve the results.
 Moreover, we note here that in contrast to the $B(E2\downarrow$) values for
 the $4^+_1 \to 2^+_1$ and $6^+_1 \to 4^+_1$ transitions with obvious minima at $\gamma=10^\circ$,
 $B(E2\downarrow$) values for $2^+_1\to 0^+_1$ changes only moderately with $\gamma$,
 ranging from $72$ e$^2$fm$^4$ to $88$ e$^2$fm$^4$, which is consistent with the data $86.4\pm1.6$
 e$^2$fm$^4$ of Ref.~\cite{Keinonen89}.

 \subsubsection{Application to $^{30}$Mg}

 The evolution of shell structure and appearance of new magic numbers
 in neutron-rich nuclei has become one of the main topics in recent
 investigations of nuclear structure physics. Especially, the erosion
 of the neutron magic numbers $N=20$ and 28 and the occurrence of well-deformed
 prolate deformed structures in such magic or close-to-magic nuclei are
 presently in the focus of several investigations.

 There is much controversy about the deformation of the ground state
 in the nucleus $^{30}$Mg. Experimentally, this deformation
 is determined by measuring the $B(E2; 0^+_{\rm gs} \rightarrow 2^+)$
 transition probability. The values obtained at MSU and at GANIL using the method of
 intermediate-energy Coulomb excitation are 295(26) e$^2$ fm$^4$~\cite{Pritychenko99} and
 435(58) e$^2$ fm$^4$~\cite{Chiste01}, respectively. However, the most recent measurement
 performed at CERN results in 241(31) e$^2$ fm$^4$~\cite{Niedermaier05}, which
 is lower than those extracted in previous measurements performed at intermediate energies.
 Therefore it is very interesting to study this problem theoretically within the present
 approach.

\begin{figure}[]
  \centering
  \includegraphics[width=4cm]{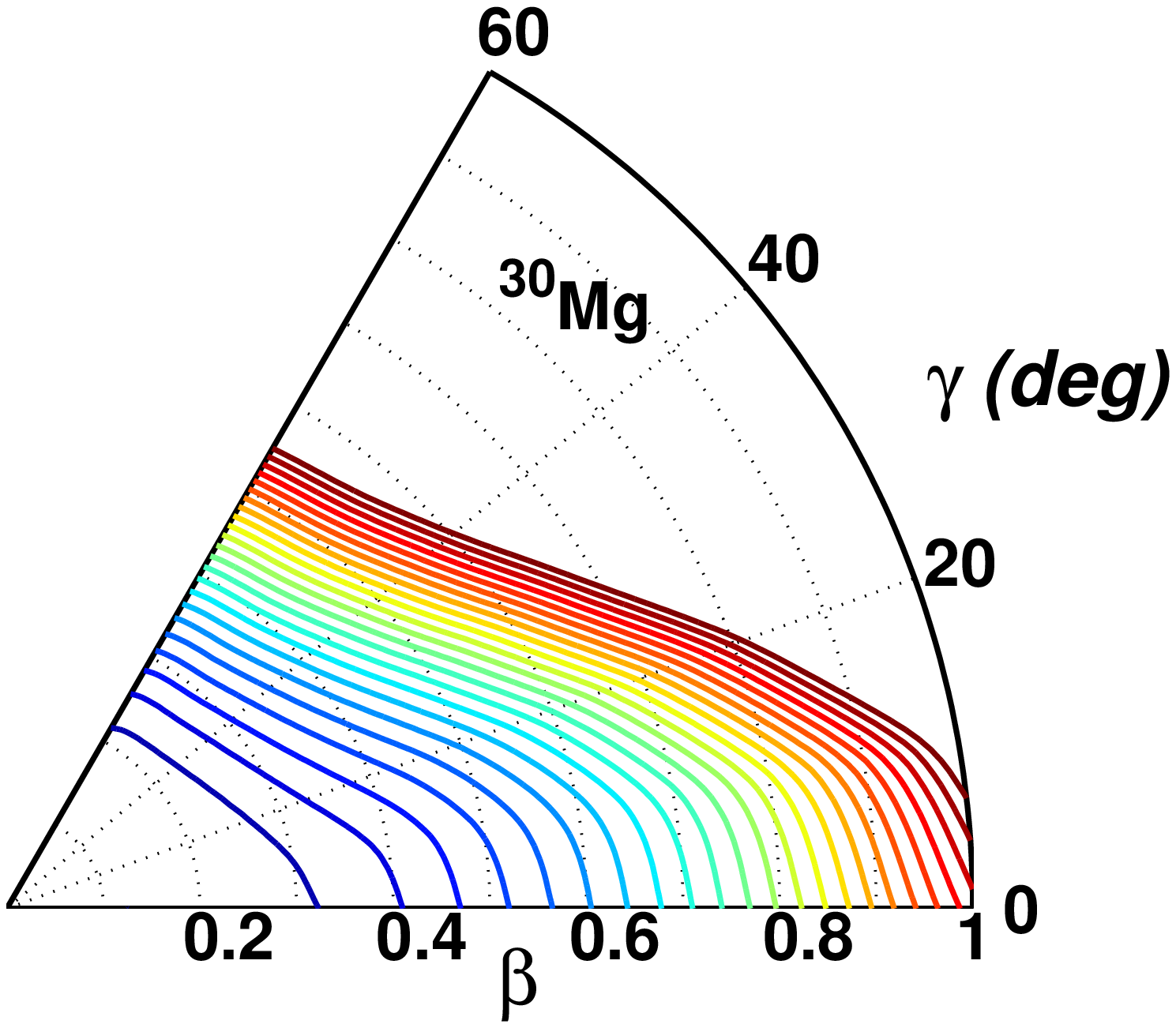}%
  \includegraphics[width=4cm]{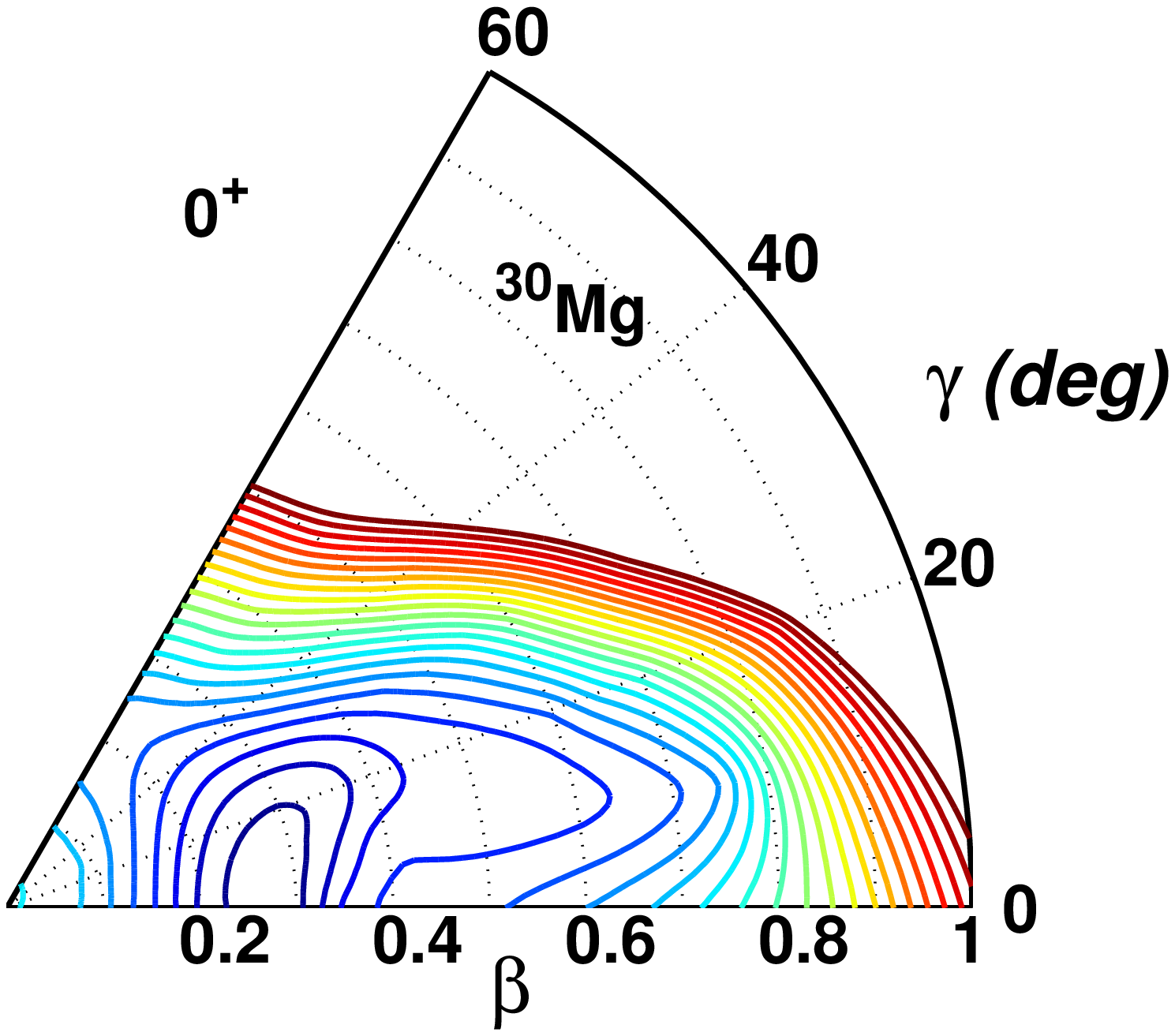}%
   \caption{(Color online) The potential energy surfaces of mean-field
   state and projected $0^+$ states in the $\beta$-$\gamma$ plane
   obtained by triaxial RMF-PC+BCS
   calculations for the nucleus $^{30}$Mg.
   The contour lines are separated by $0.5$~MeV.}
   \label{fig18}
 \end{figure}

 In Fig.~\ref{fig18} we plot the potential energy surfaces of mean-field
 states and projected $0^+$ states in the $\beta$-$\gamma$ plane for the nucleus
 $^{30}$Mg. The intrinsic states are calculated in the triaxial RMF-PC+BCS approach
 using monopole pairing forces with $G_n=24.4/A$, $G_p=29.7/A$, adjusted to the
 experimental odd-even mass differences.
 We find that the mean-field potential energy surface is very soft against $\beta$
 in the spherical region. It is hard to recognize a minimum. The
 energy surface projected on the $0^+$-state has, however, a pronounced axially symmetric
 minimum.

\begin{figure}[]
  \centering
  \includegraphics[width=9cm]{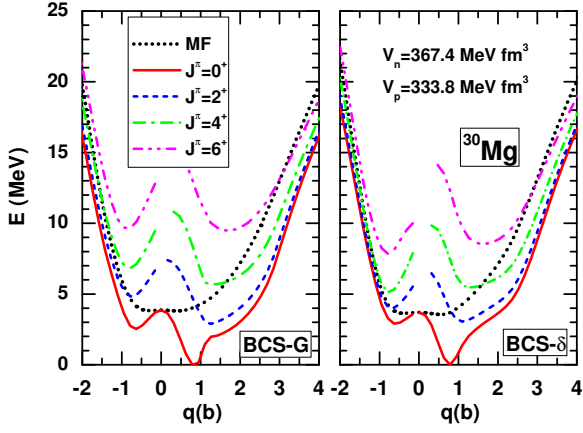}
   \caption{(Color online) Potential energy curves of the projected
   $J^\pi=0^+,2^+,4^+,6^+$ states in $^{30}$Mg, as functions of the
   quadrupole moment $q$. The intrinsic deformed states are obtained
   by RMF-PC+BCS calculations with both monopole pairing forces
   (left panel) and $\delta$-type pairing forces (right panel).
   The pairing strength parameters $V_\tau$ for the zero range pairing
   forces are adjusted the experimental pairing gap as discussed in Eq.~(\ref{avgap}).}
   \label{fig19}
 \end{figure}

 Fig.~\ref{fig19} shows axially symmetric results for the nucleus states in $^{30}$Mg.
 The corresponding potential energy curves of the intrinsic states and of the projected
 $J^\pi=0^+,2^+,4^+,6^+$ states are given as functions of the quadrupole moment $q$ ($q_{22}=0$).
 The intrinsic deformed states are obtained in the RMF-PC+BCS approach using either a monopole
 pairing forces or a zero range $\delta$-type pairing forces. We find that the projected curves
 for the $0^+$ state have in both cases an obvious minimum at $\beta\simeq0.25$.
 The energy differences between the minimum and the spherical shape are
 3.87 MeV (BCS-G) and 3.69 MeV (BCS-$\delta$) respectively. The corresponding
 $B(E2: 0^+\rightarrow 2^+)$ values are $194.8$ e$^2$fm$^4$
 and $194.6$ e$^2$fm$^4$, respectively. Both of them are somewhat
 smaller than the data.

 \subsubsection{Application to $^{32}$Mg}

 For the nucleus $^{32}$Mg, a much lower excitation energy of 0.885 MeV
 was measured for the first $2^+$-state~\cite{Mueller84}
 and a large deformation with $\beta\simeq0.51$ has been inferred
 from the measured B(E2: $0^+\rightarrow2^+$) value
 ($454\pm78$ e$^2$fm$^4$)~\cite{Motobayashi95}. Therefore this nucleus has
 drawn much attention in studies with self-consistent
 approaches. Corrections from the angular momentum projection and
 configuration mixing are found to be essential to reproduce the
 large deformed ground state of $^{32}$Mg in the HFB approach with
 the Gogny force D1S~\cite{Guzman00plb,Guzman00prc}.
 However, similar non-relativistic calculation with Skyrme-type the Sly4
 force~\cite{Heenen00} and relativistic calculation with the PC-F1 force fail
 to reproduce the data~\cite{Niksic06I}.  Therefore, it is interesting to revisit
 this problem within our 3DAMP+RMF-PC approach.
 \begin{figure}[h!]
  \centering
  \includegraphics[width=9cm]{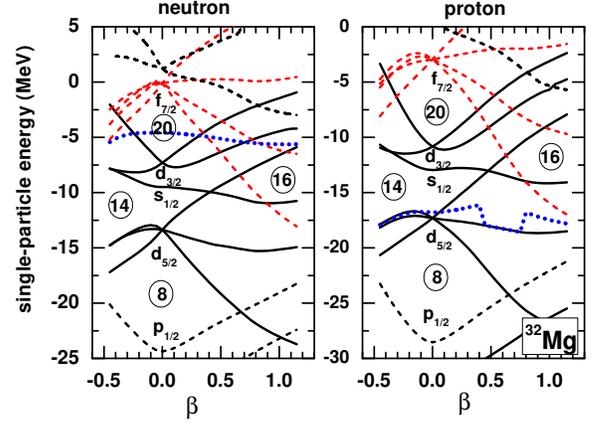}
   \caption{(Color online) The neutron (left panel) and proton (right panel)
   single-particle levels for $^{32}$Mg,
   as functions of the quadrupole deformation $\beta$.
   The levels with positive (negative) parity are shown with solid (dashed) lines.
   The levels belonging to the $f_{7/2}$ orbit are plotted with red
   dashed lines. The fermi energies for neutrons and protons are plotted
   with blue dotted lines.}
   \label{fig20}
 \end{figure}

 \begin{figure}[]
  \centering
  \includegraphics[width=8cm]{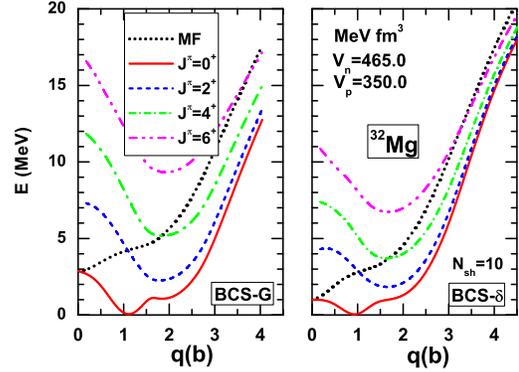}
   \caption{(Color online) Potential energy curves of the projected
   $J^\pi=0^+,2^+,4^+,6^+$ states in $^{32}$Mg, as functions of the
   quadrupole moment $q$. The intrinsic deformed states are obtained
   by RMF-PC+BCS calculations with monopole forces (left panel)
   and $\delta$-forces (right panel).
   The pairing strength parameters $V_\tau$ for the zero range pairing forces are adjusted
   the experimental pairing gap as discussed in Eq.~(\ref{avgap}).}
   \label{fig21}
 \end{figure}%

 Fig.~\ref{fig20} displays the neutron and proton RMF-PC+BCS single-particle energy levels
 for $^{32}$Mg as functions of the quadrupole deformation $\beta$. The pairing strength
 parameters are $G_n=26.78/A$ and $G_p=32.25/A$ for the monopole
 pairing force. They are obtained by adjusting the gaps at the spherical minimum
 (the ground state of the mean-field calculation) to the experimental odd-even mass
 difference with a five-point formula. In the self-consistent calculations we find
 a collapse of proton pairing for the range $0.45<\beta<0.75$.

  The potential energy curves of the projected $J^\pi=0^+,2^+,4^+,6^+$ states
  in $^{32}$Mg are plotted in Fig.~\ref{fig21} as functions of the quadrupole moment
  $q$ ($q_{22}=0$). The intrinsic deformed states are obtained from RMF-PC+BCS calculations with
  monopole forces and $\delta$-forces. The pairing strengthes $V_\tau$
  are adjusted to the odd-even mass difference.

 \begin{figure}[]
  \centering
  \includegraphics[width=6cm]{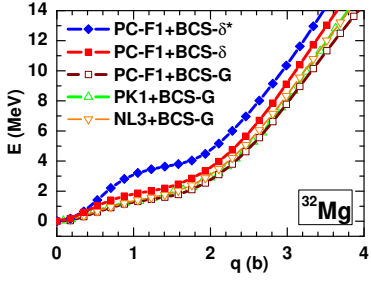}
   \caption{(Color online) The energy curves as functions of mass quadruple
   moment $q$ ($\gamma=0^\circ$) for $^{32}$Mg, calculated with
   different RMF parameterizations.}
   \label{fig22}
 \end{figure}

 At the mean-field level, a shoulder of only 1.8 MeV above the spherical minimum
 has been found in the present calculations with pairing strength parameters adjusted
 to odd-even mass differences. This value is close to the prediction of 1.9 MeV for
 the shoulder by the HFB approach with the Gogny force~\cite{Guzman02npa},
 but much smaller than the value of 3.5 MeV predicted by the RMF-PC model
 with $\delta$ pairing forces taken from the parameter set PC-F1 set~\cite{Niksic06I}.
 In Fig.~\ref{fig22} we show various RMF calculations for this shoulder with
 the parameter sets PC-F1~\cite{Niksic06I}, PK1~\cite{Long04} and
 NL3~\cite{Lalazissis97}.
 Pairing correlations are taken into account by the BCS method with
 a monopole pairing force (BCS-G) or a $\delta$-force (BCS-$\delta$).
 In all cases the pairing strength
 parameters are adjusted to the odd-even mass difference except the case labeled
 by ``BCS-$\delta$*" where $V_\tau$ has bee taken from the PC-F1 set~\cite{Niksic06I}.

 We find that the energy curves in RMF calculations do not depend
 on too much on the effective interactions but rather strongly on the
 strength of the pairing force. All the calculations with a pairing strength
 adjusted to the experimental pairing gaps give a lower shoulder,
 while the calculation with a $\delta$-pairing forces taken
 from the PC-F1 set produce a higher shoulder with a stiffer energy
 surface against quadrupole deformation $\beta$.
 Similar phenomena have also been found in Skyrme-Hartree-Fock+BCS
 calculations~\cite{Guzman07}. As a consequence, one will obtain
 different predictions for the deformation of ground state
 for different pairing correlations. More detailed investigations
 concerning this question are in progress.

 \begin{figure}[]
  \centering
  \includegraphics[width=4cm]{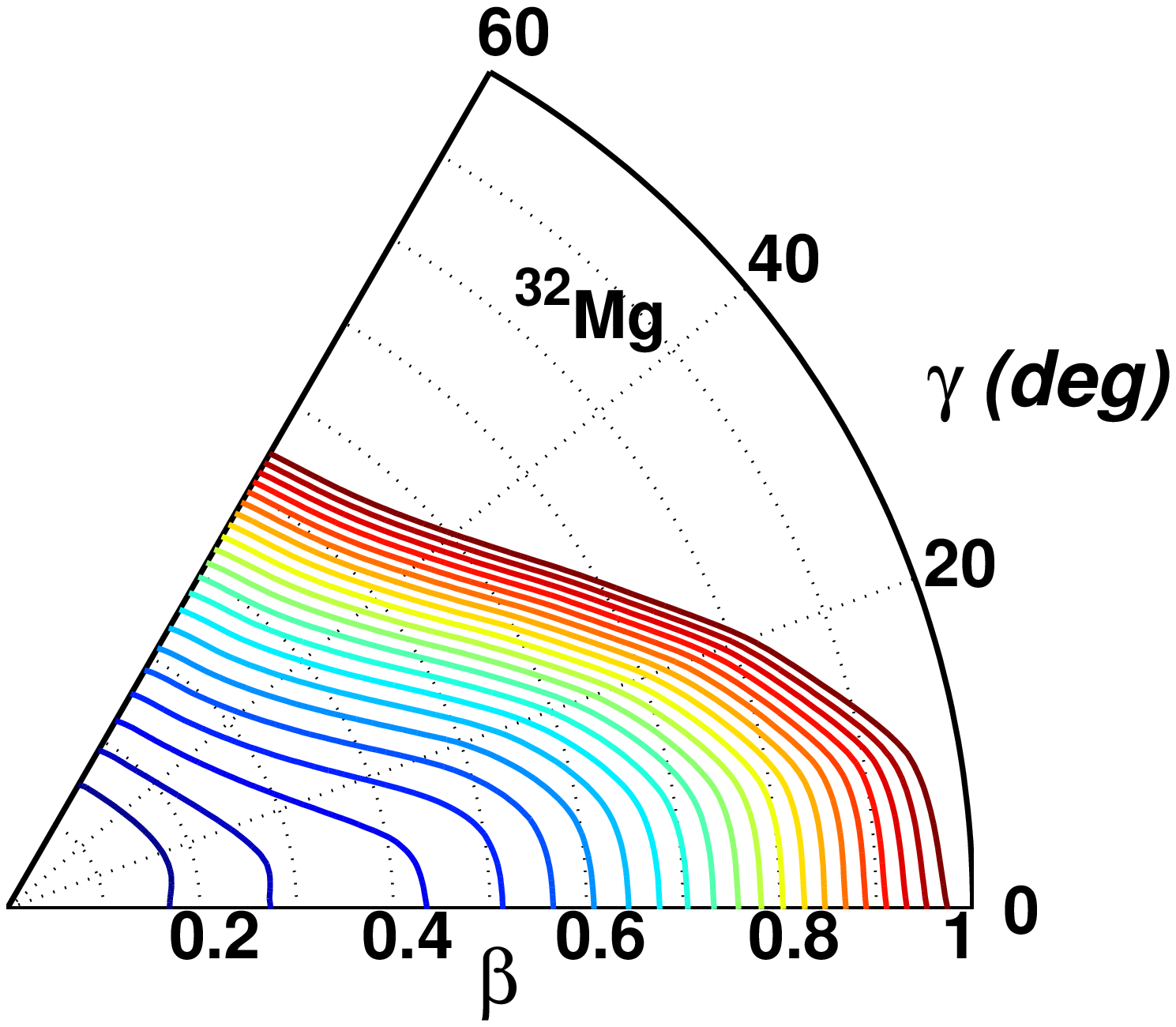}
  \includegraphics[width=4cm]{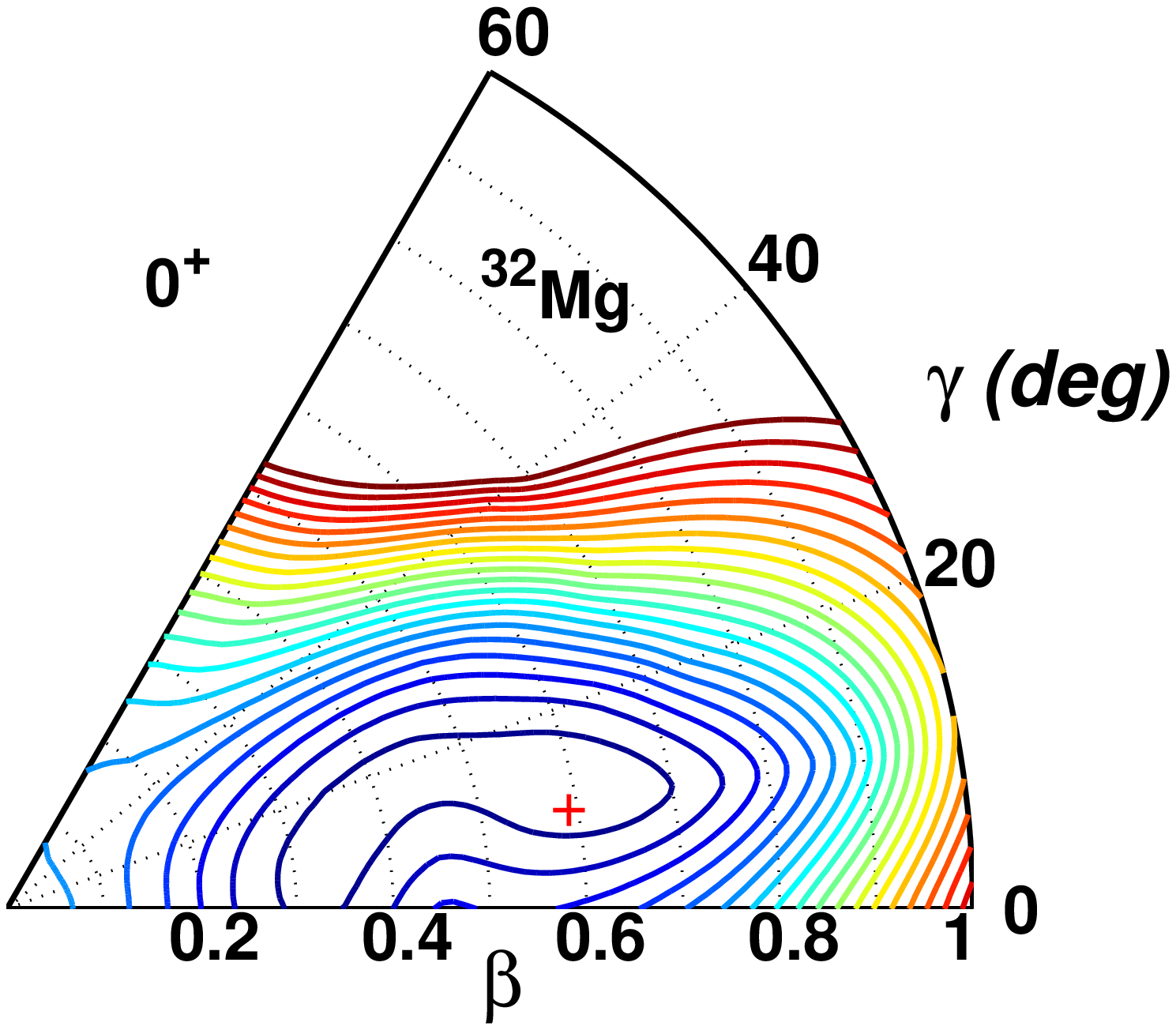}
   \caption{(Color online) The potential energy surfaces of
   mean-field theory (left panel) and of angular momentum projection $J=0$
   after the variation (right panel) in the $\beta$-$\gamma$ plane obtained
   by triaxial RMF-PC+BCS calculations for $^{32}$Mg.  The quadrupole
   deformation of the minimum
   in potential energy surface of $0^+$ state is $\beta\simeq0.6,\gamma\simeq10^\circ$.
   The contour lines are separated by $0.5$~MeV.}
   \label{fig23}
 \end{figure}

 In Fig.~\ref{fig23} we examine the potential energy surface in the $\beta$-$\gamma$ plane
 for $^{32}$Mg. Triaxial RMF-PC+BCS calculations with a monopole pairing force (left panel)
 are compared with angular momentum projection on $J=0$. We observe
 that considering the $\gamma$-degree of freedom one can expect
 considerably enlarged ground state deformations. The quadrupole deformation
 of the minimum in the angular projected $0^+$ PES is found to be at
 $\beta\simeq0.6,\gamma\simeq10^\circ$, based on which,
 the predicted energy of the $2^+$ state is $E=1.21$ MeV and the predicted
 B(E2: $0^+\rightarrow2^+$) value is $573.5$ e$^2$fm$^4$. It has to be pointed out that
 the PES of $0^+$ state is very $\beta$-soft in the region $0.3\leq\beta<0.7$.
 Based on the intrinsic state with quadrupole deformations $\beta=0.3, \gamma=0$,
 the AMP predicted an energy of $2^+$ is $E=3.39$ MeV and a B(E2: $0^+\rightarrow2^+$)
 value of $250.2$ e$^2$fm$^4$. This indicates clearly that the generator coordinate
 method based on 3DAMP approach becomes necessary for a full understanding of the
 properties of $^{32}$Mg.

 \section{Summary and perspective}
 \label{Sec.IV}

 In this paper, a full three-dimensional angular momentum projection on
 top of a triaxial relativistic mean-field calculation has been implemented
 for the first time. The underlying Lagrangian is a point coupling
 model and pairing correlations are taken into account both with a monopole
 force and a $\delta$-force. Convergence has been checked and the
 validity of this newly-developed approach has been illustrated by applying it
 to the description of the low-lying excited states in several Mg isotopes.

 For $^{24}$Mg no pronounced minimum with obvious triaxial deformation
 has been found on the potential energy surface of the $0^+$ state.
 A minimum with $\beta\approx0.55,~\gamma\approx10^\circ$ has been
 found on the PES of the first $2^{+}$ state. Using this minimum
 as a basis for the projection the
 experimentally observed excitation energies and B(E2) transition
 probabilities can be qualitatively reproduced. However, the
 predicted spacing between the levels is overestimated in this
 approach.

 For $^{30}$Mg, the projected energy surface of the $0^+$ state
 has a obvious minimum with $\beta\simeq0.25$. The energy differences between
 the minimum and the spherical shape are
 3.87 MeV (BCS-G) and 3.69 MeV (BCS-$\delta$) respectively. The corresponding
 $B(E2: 0^+\rightarrow 2^+)$ are respectively $194.8$ e$^2$fm$^4$
 and $194.6$ e$^2$fm$^4$.

 For $^{32}$Mg, we note that the calculations with adjusted pairing
 strength parameters produce always a lower shoulder in the mean-field energy
 curve, which is, together with the triaxial degree of freedom, essential to
 reproduce the large deformed ground state.
 Moreover, the mean-field and the projected $0^+$ potential energy surfaces of
 $^{32}$Mg have been found to be very $\gamma$-soft in the region of small deformations
 and $\beta$-soft in the neighborhood of its minimum.

 These investigations indicate that, besides triaxiality, the effects of pairing
 correlations and shape fluctuations should be treated more carefully in the
 description of low-lying excited states of exotic nuclei. Work in this direction is in
 progress.

 Finally, we would like to point out that the pairing strength parameters of
 protons and neutrons in PC-F1 are adjusted to the pairing gaps of the nuclei:
 $^{136}$Xe, $^{144}$Sm, $^{112}$Sn, $^{120}$Sn and $^{124}$Sn respectively.
 However, pairing strength parameter obtained in this way might not be well-justified
 in the region of light nuclei. We have found in the present study for $^{30}$Mg:
 $f_{p}=1.04$, $f_{n}=1.19$ and for $^{32}$Mg: $f_{p}=1.51$,
 $f_{n}=1.09$, where $f_{\tau}$ is the ratio of the pairing strength
 parameters of the adjusted delta-pairing and the standard PC-F1
 delta-pairing. It indicates that a better parameterizations of the
 energy density functional is required for the description of light
 nuclei.


 \begin{acknowledgments}
 Helpful discussions with D. Vretenar are gratefully acknowledged.
This research has been supported by the Asia-Europe Link Project
[CN/ASIA-LINK/008 (094-791)] of the European Commission, the
National Natural Science Foundation of China under Grant No.
10775004, 10221003, 10720003, 10705004, the Bundesministerium
f\"{u}r Bildung und Forschung, Germany under project 06 MT 246 and
by the DFG cluster of excellence \textquotedblleft Origin and
Structure of the Universe\textquotedblright\
(www.universe-cluster.de).
 \end{acknowledgments}

 \begin{appendix}

 \section{Evaluation of contractions and overlaps}
\label{AppendixA}
 The contractions and overlaps have been derived in detail in Ref.~\cite{Valor00},
 where, however, the rotation matrix is assumed to be real from the beginning.
 This is no longer the case for a three-dimensional angular momentum
 projection. In addition these earlier investigations were done only
 for nonrelativistic density functionals. Therefore, we derive here
 in a similar way the general formulae of the contractions and overlaps
 suitable for the three-dimensional relativistic case and used in
 the present numerical applications.

  \subsection{Determination of the generalized contractions}
 In the following we derive formulae of generalized contractions
 $\langle \Phi(q_a)\vert\hat O\hat R(\Omega)\vert\Phi(q_b)\rangle$
 connecting different intrinsic states. Such formulae can be applied directly
 in future Generator Coordinate (GCM) calculations with 3DAMP as well.

 For convenience, we introduce the following notation
 \beq
 \label{E36}
 {}_a\langle 0 \vert    \equiv \langle\Phi(q_a)\vert ,\quad
 \vert \Omega \rangle_b \equiv \frac{\hat R(\Omega)\vert \Phi(q_b)\rangle}
                             {\langle \Phi(q_a)\vert\hat R(\Omega)\vert \Phi(q_b)\rangle}.
 \eeq
 The quasiparticle vacua $\vert 0 \rangle_a$ and $\vert \Omega \rangle_b$
 are defined by the corresponding quasiparticle operators $\alpha_k$ and $\beta_k$ respectively,
 \beq
 \label{HFB-appro}
  {}_a\langle 0\vert \alpha^\dagger_k =0,\quad \beta_k\vert \Omega \rangle_b=0.
 \eeq
 According to the generalized Wick theorem the contractions
 $\langle \Phi(q_a)\vert\hat O\hat R(\Omega)\vert\Phi(q_b)\rangle$
 for an arbitrary many-body operator $\hat O$ can be expressed in
 terms of the mixed densities and mixed pairing tensors
 \bsub
 \label{mixed}
 \beqn
 \label{Mixdensmat}
  \rho_{kl} (q_a,q_b;\Omega)
  &\equiv& {}_a\langle 0\vert a^\dagger_la_k \vert \Omega\rangle_b,\\
 \label{Mixpaimat1}
   \kappa^{10}_{kl} (q_a,q_b;\Omega)
   &\equiv& {}_a\langle0\vert  a_la_k \vert \Omega\rangle_b,\\
 \label{Mixpaimat2}
    \kappa^{01}_{kl} (q_a,q_b;\Omega)
    &\equiv& {}_a\langle0\vert  a^\dagger_ka^\dagger_l \vert \Omega\rangle^\ast_b.
  \eeqn
 \esub
In order to derive expressions for these mixed densities we consider
the fact that the quasiparticle operators ($\alpha,\alpha^\dagger$)
and ($\beta,\beta^\dagger$) are connected by a Bogoliubov
transformation~\cite{Ring80}
 \beq
 \label{Ttransformation}
 \begin{pmatrix}
    \alpha\\
    \alpha^\dagger
 \end{pmatrix} =
 \begin{pmatrix}
    \mathbb{U}^\dagger & \mathbb{V}^\dagger \\
    \mathbb{V}^T       & \mathbb{U}^T
 \end{pmatrix}
\begin{pmatrix}
    \beta\\
    \beta^\dagger
 \end{pmatrix}.
 \eeq
 On the other hand, the quasiparticle operators $\alpha, \alpha^\dagger$
 are related to the particle operators $a, a^\dagger$ by a Bogoliubov
 transformation,
\beq
 \label{Aquasip}
 \begin{pmatrix}
    \alpha\\
    \alpha^\dagger
\end{pmatrix} =
 \begin{pmatrix}
    U^\dagger_a & V^\dagger_a \\
    V^T_a       & U^T_a
 \end{pmatrix}
 \begin{pmatrix}
    a         \\
    a^\dagger
\end{pmatrix}.
\eeq%
In a similar way the quasiparticle operators $\beta, \beta^\dagger$
 are related to the particle operators $b, b^\dagger$ by
\beq
 \label{EA6}
 \begin{pmatrix}
    \beta \\
    \beta^\dagger
 \end{pmatrix} =
 \begin{pmatrix}
    U^\dagger_b & V^\dagger_b \\
    V^T_b       & U^T_b
 \end{pmatrix}
 \begin{pmatrix}
    b        \\
    b^\dagger
 \end{pmatrix}.
 \eeq
 Assuming that the operators $a, a^\dagger$
 and $b, b^\dagger$ are related by a rotation as~\cite{Robledo94}
 \beq
  \begin{pmatrix}
         a        \\
         a^\dagger
  \end{pmatrix}=
  \begin{pmatrix}
         R(\Omega) & 0     \\
         0 & R^\ast(\Omega)
  \end{pmatrix}
  \begin{pmatrix}
        b        \\
        b^\dagger
  \end{pmatrix},
 \eeq
 one finds for the particle operators $a,a^\dagger$ and
 the quasiparticle operators $\beta,\beta^\dagger$ the relation
 \beqn
 \label{Bquasip}
      \begin{pmatrix}
          \beta         \\
          \beta^\dagger
      \end{pmatrix}
     =\begin{pmatrix}
                U^\dagger_b(\Omega) & V^\dagger_b(\Omega) \\
                V^T_b(\Omega)       & U^T_b(\Omega)
       \end{pmatrix}
       \begin{pmatrix}
                a        \\
                a^\dagger
       \end{pmatrix},
 \eeqn
 with the coefficients $U_b(\Omega),V_b(\Omega)$ given by
  \beq
  U_b(\Omega) = R(\Omega) U_b,\quad
  V_b(\Omega) = R^\ast(\Omega) V_b.
  \eeq
Combining Eq.~(\ref{Aquasip}) and Eq.~(\ref{Bquasip}) we obtain the
matrices $\mathbb{U}$ and $\mathbb{V}$ in Eq.~(\ref{Ttransformation})
\bsub%
\beqn%
 \label{Tmatrix}
  \mathbb{U}^\dagger  &=& U^\dagger_a R(\Omega)U_b + V^\dagger_a R^\ast(\Omega) V_b,\\
  \mathbb{V}^\dagger &=& U^\dagger_aR(\Omega)V^\ast_b+V^\dagger_a R^\ast(\Omega)
  U^\ast_b,
\eeqn%
\esub%
 that relates the quasiparticle operators $\beta,\beta^\dagger$ and
 $\alpha,\alpha^\dagger$ and the
 quasiparticle vacua ${}_a\langle 0 \vert$ and
 $\vert \Omega \rangle_b$ in Eq.~(\ref{E36}).
 With the help of generalized Wick's
 theorem~\cite{Balian69}, one finds the contraction
 \beq
 {}_a\langle 0\vert \alpha \beta^\dagger \vert\Omega\rangle_b
 =\mathbb{U}^{-1},
 \eeq
 and in combination with Eqs.~(\ref{HFB-appro}), (\ref{Aquasip}) and (\ref{Bquasip}),
 the elements of the mixed density and the mixed pairing tensors of Eq.~(\ref{mixed})
 are obtained as
 \bsub
 \label{mixedx}
 \beqn
 \label{Mixdensmatx1}
  \rho_{kl} (q_a,q_b;\Omega)
  &=& [V^\ast_b(\Omega) [\mathbb{U}^{T}]^{-1} V^T_a]_{kl},\\
 \label{Mixpaimatx1}
   \kappa^{10}_{kl} (q_a,q_b;\Omega)
   & = & [V^{\ast}_b(\Omega) [\mathbb{U}^{T}]^{-1} U^T_a]_{kl},\\
 \label{Mixpaimatx2}
    \kappa^{01}_{kl} (q_a,q_b;\Omega)
   & = & [U^\ast_b(\Omega) [\mathbb{U}^{T}]^{-1} V^T_a]^\ast_{lk}.
  \eeqn
 \esub

\subsection{Restriction to the occupied space}

 In practical three-dimensional applications the matrices $U$, $V$,
 $\mathbb{V}$ etc. have the very large dimension of
 the oscillator basis. In fact most of the high-lying eigenstates
 of the Dirac equation are not occupied and therefore they do not contribute to the overlap
 integrals. In order to reduce the computational effort it is
 therefore of great importance to eliminate these high-lying
 eigenstates in the Dirac basis, where the mean field wave function has
 the form of a BCS wave function. The procedure
 discussed in the following is, however, not restricted to
 RMF+BCS calculations used in this investigation. In general
 Hartree-Bogoliubov theory one can apply
 similar formulae in the canonical basis~\cite{Bloch62,Ring80} where an
 arbitrary Hartree-Bogoliubov wave function has BCS form.
 In this basis the
 intrinsic states $\vert\Phi(q)\rangle$ are characterized by the
 special Bogoliubov-Valatin transformation of the form
 \beqn
  \label{BCS-transfor}
  \bar U = \begin{pmatrix}
             u_k & 0\\
              0  & u_k
      \end{pmatrix},\quad
  \bar V = \begin{pmatrix}
            0 & v_k\\
         -v_k & 0
     \end{pmatrix}.
  \eeqn
 Here $u_k, v_k$ are real positive numbers and the phase has been chosen as
 $u_{\bar k}=u_k$, $v_{\bar k}=-v_k$, where $\bar k$ is the time reversed state
 of $k$.
 Since unoccupied states with $v^2_k=0$ have no contribution to overlap
 and contractions, one can eliminate these states to simplify the
 calculation~\cite{Bonche90,Robledo94,Valor00}. As usual, one can introduce
 a cut-off $\zeta$ to divide the full Dirac space into two parts:
 an occupied part with $v^2_k>\zeta$ and an unoccupied part
 with $v^2_k\leq\zeta$ and the matrices $U$ and $V$ in Eqs.~(\ref{Aquasip})
 and (\ref{EA6}) have the form
 \beqn
 \label{Division}
  V =  \begin{pmatrix}
            \bar V & 0\\
                0  & 0
       \end{pmatrix},\
 U =   \begin{pmatrix}
            \bar U & 0\\
                0  & 1
       \end{pmatrix},\
 R =   \begin{pmatrix}
             {\bar R} & R_{10}\\
               R_{01} & R_{00}
       \end{pmatrix}.~~~~~~~
 \eeqn
 The matrix ${\bar R}$ is related to the occupied states only.
 In this case the matrix $\mathbb{U}^T$ in Eq.~(\ref{Tmatrix}) becomes
 \beqn
 \mathbb{U}^T = \begin{pmatrix}
 \bar U^T_a{\bar R}^\ast\bar U^\ast_b + \bar V^T_a{\bar R} \bar V^\ast_b
                  & \bar U^T_a R^\ast_{10}\\
  R_{01}^\ast\bar U^\ast_b           & R_{00}^\ast
           \end{pmatrix}
 \eeqn
 and its inverse has the form
\begin{widetext}
\beq
\label{T22inv}
  [\mathbb{U}^{T}]^{-1}
 = \left(
 \begin{array}[c]{cc}%
    {\bar D}^{-1}  & {\bar D}^{-1}\bar U^T_a({\bar R}^T)^{-1}{R^T_{01}}    \\
    {R^T_{10}}({\bar R}^T)^{-1}\bar U^\ast_b{\bar D}^{-1}; &  (R_{00}^\ast)^{-1}+
   {R^T_{10}}({\bar R}^T)^{-1}\bar U^\ast_b {\bar D}^{-1}
   {\bar U}^T_a ({\bar R}^T)^{-1}R^T_{01}
   \end{array}\right),
\eeq
\end{widetext}
 where the matrix ${\bar D}$ is defined as,
 \beq
 \label{Dmatrix}
  {\bar D} =\bar U^T_a({\bar R}^T)^{-1}\bar U^\ast_b +\bar V^T_a{\bar R}\bar V^\ast_b.
 \eeq
In the general case of GCM calculations where $q_a \ne q_b$ the
BCS-space of the wave function $\vert \Phi(q_a)\rangle$ is different
from the BCS-space of the wave function $\vert \Phi(q_b)\rangle$ and
therefore the cut-off procedure can lead to occupied subspaces and to
matrices $\bar V_a$ and $\bar V_b$ with different dimensions and
rectangular matrices $\bar R$ and $\bar D$, which cannot be inverted.
In such cases appropriate cut-off parameters $\zeta_a$ and $\zeta_b$
have to be chosen, such that the matrix $\bar D$ stays a square
matrix.

The elements of the mixed density in Eq.~(\ref{mixedx}) are%
\beq
\rho_{kl}(q_{a},q_{b};\Omega)  = \left(
\begin{array}
[c]{cc}%
\bar{R}\bar{V}_{b}^{\ast}{\bar{D}}^{-1}\bar{V}_{a}^{T} & 0\\
R_{01}\bar{V}_{b}^{\ast}{\bar{D}}^{-1}\bar{V}_{a}^{T} & 0
\end{array}
\right).
\eeq%
The matrices  $R_{01}$ and $R_{10}$ connect the occupied space with
the unoccupied space by rotation. We neglect these matrices in the
mixed densities and pairing tensors, because they are usually very
small, i.e. we restrict ourselves to the occupied space in the
further calculations. In this space we obtain the elements of the
mixed density and the mixed pairing tensors as:
 \bsub
 \label{E50}
 \beqn
 \label{mixdensmatBCS}
  \bar \rho_{kl} (q_a,q_b;\Omega)
 &=& [{\bar R}\bar V^\ast_b {\bar D}^{-1}\bar V^T_a]_{kl},\\
 \label{mixpairmatBCS1}
 \bar \kappa^{10}_{kl} (q_a,q_b;\Omega)
 &=& [{\bar R}\bar V^\ast_b {\bar D}^{-1}\bar U^T_a]_{kl},\\
 \label{mixpairmatBCS2}
 \bar \kappa^{01}_{kl}(q_a,q_b;\Omega)
 &=& [{\bar R}^\ast\bar U^\ast_b {\bar D}^{-1}\bar V^T_a]^\ast_{lk}.
 \eeqn
 \esub
This shows that we finally have to invert only the matrix ${\bar D}$
in the occupied subspace. The explicit expressions for the matrix
elements of ${\bar D}$
 in Eq.~(\ref{Dmatrix}) are
 \bsub \beqn
 {\bar D}_{kl}           &=& u^a_k ({\bar R}^T)^{-1}_{kl}u^b_l + v^a_k {\bar R}^\ast_{kl}v^b_l,\\
 {\bar D}_{k\bar l}      &=& u^a_k ({\bar R}^T)^{-1}_{k\bar l}u^b_l + v^a_k {\bar R}^\ast_{k\bar l}v^b_l,
 \eeqn\esub
 where the indices $k,l$ run over the states with non-vanishing
 occupation numbers. Using the time reversal properties of the rotational
 operator, one finds the following relations:
 \beq
 {\bar D}_{\bar kl}       = -{\bar D}^\ast_{k\bar l},\quad
 {\bar D}_{\bar k\bar l}  = {\bar D}^\ast_{kl}.
 \eeq

\subsection{Determination of the overlaps}

 The norm overlap has already been derived in Ref.~\cite{Balian69},
 \beq
  \label{Norm-OVLAP}
   \langle \Phi(q_a)\vert \hat R(\Omega)\vert \Phi(q_b)\rangle
    =\pm\sqrt{\det \mathbb{U}}.
 \eeq
After some calculations we obtain%
\begin{widetext}
\beq \mathbb{U}^{T} =
U_{a}^{T}{R}^{\ast}U_{b}^{\ast}+V_{a}^{T}{R}V_{b}^{\ast}
=\left(\begin{array}%
[c]{cc}%
1 &
\bar{V}_{a}^{T}\bar{R}\bar{V}_{b}^{\ast}\bar{U}_{b}^{\ast-1}{R}^T_{01}\\
 {0} & 1
\end{array}\right)%
\left(\begin{array}%
[c]{cc}%
\bar{U}_{a}^{T}\bar{R}^{T-1}\bar{U}_{b}^{\ast}+\bar{V}_{a}^{T}\bar{R}\bar
{V}_{b}^{\ast} & 0\\
{0} & 1
\end{array}\right)%
\left(\begin{array}
[c]{cc}%
\bar{U}_{b}^{\ast-1}\bar{R}^{T}\bar{U}_{a}^{-1T} & 0\\
{0} & 1
\end{array}\right)%
U_{a}^{T}{R}^{\ast}U_{b}^{\ast}%
\eeq%
\end{widetext}%
 and using $\det R =1$ we find that the norm overlap in
Eq.~(\ref{Norm-OVLAP}) is simply a product of two determinants of
much smaller dimension:%
\beq \det\mathbb{U}=\det{\bar D}\det{\bar R}.%
\eeq%
The phase of the overlap in Eq.~(\ref{Norm-OVLAP}) remains open.
Neerg{\aa}rd and W\"ust~\cite{Neergard83} pointed out that the phase
 problem of the norm overlap could be avoided by rewriting the norm
 overlap into the following form
 \beqn
 \label{Neegard}
  \langle \Phi(q_a)\vert \hat R(\Omega)\vert \Phi(q_b)\rangle
 &=&(\prod_{k>0}u_ku^\prime_k)\sqrt{\det[1+M]}~~~~~~~~~\nonumber\\
 &=&(\prod_{k>0}u_ku^\prime_k)\prod_{l>0}(1+c_l),
 \eeqn
 where $u_k$ and $u^\prime_k$ are the Bogoliubov-Valatin
 transformations coefficients in (\ref{BCS-transfor})
 for the intrinsic states $ \vert\Phi(q_a)\rangle$
 and $ \vert\Phi(q_b)\rangle$ respectively.
 The product $\prod_{l>0}$ runs over the pairwise degenerate
 eigenvalues $c_l$ of the matrix $M$~\cite{Beck70}
 \beq
 M(q_a,q_b;\Omega)=Z^{}_b(\Omega) Z^\dagger_a,\quad {\rm with}
 \quad Z = V^\ast U^{\ast-1}
 \eeq
 In the canonical basis the matrix $Z$ is reduced to $2\times2$-matrices of the form
 \beq
{\bar Z_k} = \begin{pmatrix}
     0               & \dfrac{v_k}{u_k}\\
    -\dfrac{v_k}{u_k} & 0 \\
  \end{pmatrix},
 \eeq
 where $k$ runs only over states with $v^2_k\geq\zeta$. In cases,
 where some of the numbers  $u_k$ vanish one can, in analogy to
 Eq.~(\ref{Division}), reduce the space intro three subspaces of fully occupied
 state ($v^2_k=1$), partially occupied states ($0<v^2_k<1$) and empty states
 ($v^2_k=0$). Finally, the norm overlap can be evaluated according to
 Eq.~(\ref{Neegard})
 by diagonalizing the matrix $M$. This method is certainly rather
 complicated. It turns out that we do not need to apply it in the
 present applications based on time reversal symmetric wave functions
 $|\Phi(q)\rangle$. The norm-overlap is 1 for $\Omega=0$, it stays
 real and positive for all values of the Euler angles
 $\Omega=(\phi,\theta,\psi)$ and therefore we have for the
 norm overlap
 \beq
  \label{Norm}
   \langle \Phi(q_a)\vert \hat R(\Omega)\vert \Phi(q_b)\rangle
    = \sqrt{\det{\bar D}\det{\bar R}}.
 \eeq

 \section{Representation of rotations in the Dirac basis}
 \label{AppendixB}

 In our calculations, the single-particle
 wave functions $\psi_{k}$ are Dirac spinors. For the solution of the
 Dirac equation
 the large and small components $f(\bm{r},s)$ and $g(\bm{r},s)$ of a Dirac
 spinor are expanded in terms of the
 eigenfunctions of a three-dimensional harmonic oscillator in Cartesian
 coordinates~\cite{Koepf88}
 \beqn
    \vert\psi_k\rangle=
     \left(
     \begin{array}{l}
      \displaystyle
      \sum\limits_{n}f_{nk} \vert n\rangle\\
      i\sum\limits_{\bar n}g_{\bar nk} \vert \bar n\rangle
     \end{array}
   \right)\chi_{t_k}(t),
 \eeqn
 where $\chi_{t_k}(t)$ is the isospin part. The harmonic
 oscillator basis states $\vert n\rangle=\vert n_x,n_y,n_z,n_s\rangle$
 with simplex $n_s=+i$ and the time reversed states $\vert \bar n\rangle$
 with simplex $n_s=-i$ are defined by
 \bsub
  \label{Basis}
  \beqn
  \vert n\rangle &=& \phi_{n_x}(x)\phi_{n_y}(y)\phi_{n_z}(z)
           \displaystyle \frac{i^{n_y}}{\sqrt{2}}%
             \left(\begin{array}{c}
                           1 \\  (-1)^{n_x+1}
             \end{array}\right),~~~~~~~~~~~~\\
   \vert\bar n\rangle &=& \phi_{n_x}(x)\phi_{n_y}(y)\phi_{n_z}(z)%
   \displaystyle\frac{(-i)^{n_y}}{\sqrt{2}}%
       \left(\begin{array}{c}
        (-1)^{n_x+1} \\   -1
       \end{array}\right),%
   \eeqn
  \esub
 where the phase factor $i^{n_y}$ is consistent with the triaxial
 self-consistent symmetries and leads to real matrix elements
 for Dirac equation~\cite{Koepf88,Yao06,PMR.08}.

 The matrix elements of the rotation operator $\hat R(\Omega)$
 in Eq.~(\ref{Division}) in the Dirac basis are derived from
 the representation of this
 operator in the harmonic oscillator basis~(\ref{Basis}) by
 \beqn
 \label{Rmatrix}
 {\bar R}_{kl}(q_a,q_b;\Omega)
 &=&\int d^3r \psi^\dagger_k(\bm{r},q_a) \hat R (\Omega) \psi_l(\bm{r},q_b)\nonumber\\
 &=&\sum_{n,n^\prime}f^\ast_{nk}(q_a)f^{}_{n^\prime l}(q_b)
       \langle n\vert\hat R(\Omega) \vert n^\prime\rangle\\
  &&+\sum_{\bar n,{\bar n}^\prime}g^\ast_{{\bar n}k}(q_a)g^{}_{{\bar n}^\prime l}(q_b)
        \langle {\bar n}\vert\hat R(\Omega) \vert{\bar n}^\prime
        \rangle.\nonumber
 \eeqn
 The rotation matrices $\langle n_1\vert\hat R(\Omega) \vert n_2\rangle$
 in the cartesian basis have been derived using the method of generating functions
 in Ref.~\cite{Nazmitdinov96}. In present work, however, we adopt a simple method
 to evaluate these matrix elements by transforming from the cartesian basis to
 the spherical oscillator basis given by $\vert m\rangle=\vert
 n_rljm\rangle$ with
 \beq
 \label{TransCoeff}
 \langle m \vert n \rangle
 =\sum_{m_l m_s} C^{jm}_{l,m_l,1/2,m_s}
 \langle n_r l m_l\vert n_xn_yn_z\rangle\langle m_s\vert n_s\rangle,
 \eeq
 where $C^{jm}_{l,m_l,1/2,m_s}$ is the Clebsch-Gordon
 coefficient. The transformation coefficients
 $\langle n_r l m_l\vert n_xn_yn_z\rangle$ are given in
 Refs.~\cite{CW.67,Talman70}. Therefore, the large and small components
 $f(\bm{r},s)$ and $g(\bm{r},s)$ of $\psi_k$ can be rewritten in terms of the
 eigenfunctions of spherical harmonic oscillator as,
 \beqn
 \vert\psi_k\rangle =\left(\begin{array}{c}
   \displaystyle
   \sum_{m}F_{mk}\vert m\rangle\\
   \displaystyle
   i\sum_{\bar m}G_{{\bar m}k}\vert\bar m\rangle
  \end{array}
  \right),
 \eeqn
 where the expansion coefficients $F_{km}$ and $G_{k \bar m}$ can
 be obtained with the help of relation in Eq.~(\ref{TransCoeff}),
 \beqn
 F_{mk} = \sum_{n}f_{nk}\langle m\vert n\rangle,\quad
 G_{{\bar m}k} = \sum_{\bar n}g_{{\bar n}k}\langle \bar m\vert\bar  n\rangle.
 \eeqn
 The matrix elements of ${\bar R}$ in Eq.~(\ref{Rmatrix}) are subsequently given by
 \beqn
 \label{B7}
 {\bar R}_{kl}(q_a,q_b;\Omega)
 &=&\sum_{mm^\prime}F^\ast_{mk}(q_a) F_{m^\prime l}(q_b)
    \langle m\vert\hat R(\Omega)\vert m^\prime\rangle\nonumber\\
 & &
  +\sum_{\bar m\bar m^\prime}G^\ast_{{\bar m}k}(q_a)G_{{\bar m}^\prime l}(q_b)
   \langle\bar  m\vert\hat R(\Omega)\vert\bar  m^\prime\rangle,\nonumber\\
 \eeqn
 where the matrix
 \beq
 \langle m\vert\hat R(\Omega)\vert m^\prime\rangle=%
  \delta_{n_rn_r^\prime}\delta_{ll^\prime}\delta_{jj^\prime}%
  D^j_{mm^\prime}(\Omega)~~~~~~~~~~
 \eeq%
 is diagonal in the quantum numbers $n_r$, $l$, $j$ and is simply given by
 the Wigner D-function. We use Condon-Shortly notation for the spherical
 harmonics $Y_{lm}(\theta,\phi)$~\cite{Edmonds57}. With the
 time reversal operator
 \beq
\vert \bar m \rangle = \hat T\vert n_rljm\rangle
 =(-1)^{l+j-m}\vert n_rlj-m\rangle,
 \eeq
 one finds the expansion coefficients of the Dirac spinor for the time reversed
state,
 \beq
  F_{m \bar k}
  =(-1)^{l+j+m}F_{-mk},\quad
  G_{m \bar k}
  =(-1)^{l+j+m+1}G_{-mk},\\
 \eeq
 where $\vert -m\rangle=\vert n_rlj-m\rangle$ and where
 $\bar k$ is the time reversed state of $k$. With these relations,
 the matrix element, ${\bar R}_{k\bar l}$ can be easily calculated.

 Moreover, according to the time reversal properties of the rotational
 operator $\hat R(\Omega)$, one immediately finds:
\beq {\bar R}_{\bar kl}=-{\bar R}^\ast_{k\bar l},\quad%
{\bar R}_{\bar k\bar l}={\bar R}^\ast_{kl}. \eeq


\section{Mixed densities in coordinate space}

 In a point coupling model with a local interaction of zero range the
 overlap integrals for the Hamiltonian are most easily evaluated in
 coordinate space. We therefore have to calculate the mixed local densities
 and currents in $r$-space. Expressing the Dirac spinors in terms of spherical
 harmonic oscillator states
\beq%
\psi_{k}(\mathbf{r)=}\left(
\begin{array}
[c]{c}%
F_{k}(\mathbf{r},\sigma)\\
iG_{k}(\mathbf{r},\sigma)
\end{array}
\right),%
\eeq with the large and small components%
\bsub%
\beqn%
F_{k}(\mathbf{r},\sigma)  & =& \sum_{m}
F_{mk}\Phi_{m}(\mathbf{r},\sigma),\\
G_{k}(\mathbf{r},\sigma)  & =&%
\sum_{m} G_{mk}\Phi_{m}(\mathbf{r},\sigma),%
\eeqn%
\esub%
and the spherical oscillator functions%
\beq \Phi_m(\bm{r},\sigma)
       =\sum_{m_l m_s}C^{jm_j}_{lm_l\frac{1}{2}m_s}R_{n_rl}(r)Y_{lm_l}(\theta,\varphi)\chi^\sigma_{m_s}.
\eeq%
Here $\chi^\sigma_{m_s}$ is the spin part.

According to Eqs.~(\ref{BCS-transfor}) and (\ref{mixdensmatBCS}) we
obtain the relativistic mixed single-particle density matrix in the
harmonic oscillator basis,
\bsub%
\label{rho++}
\beqn%
\rho^{++}_{mm^\prime}
  &=&\left[\tilde F^b(\Omega){\bar V}^{b\ast}{\bar D}^{-1}{\bar V}^{aT}F^{aT}\right]_{mm^\prime}\\%
\rho^{+-}_{m{\bar m}^\prime}
  &=&\left[\tilde F^b(\Omega){\bar V}^{b\ast}{\bar D}^{-1}{\bar V}^{aT}G^{aT}\right]_{m{\bar m}^\prime}\\%
\rho^{-+}_{{\bar m}m^\prime}
  &=&\left[\tilde G^b(\Omega){\bar V}^{b\ast}{\bar D}^{-1}{\bar V}^{aT}F^{aT}\right]_{{\bar m}m^\prime}\\%
\rho^{--}_{\bar m\bar m^\prime}
  &=&\left[\tilde G^b(\Omega){\bar V}^{b\ast}{\bar D}^{-1}{\bar V}^{aT}G^{aT}\right]_{\bar m\bar m^\prime}%
\eeqn%
\esub%
where the rotated large and small components of Dirac spinor,
$\tilde F_{mk}$
 and $\tilde G_{mk}$ are given by
 \bsub\beqn
 \tilde F_{mk}(\Omega)
 &=& \displaystyle \sum_{m^{\prime}} R_{mm^{\prime}}(\Omega) F_{m^{\prime}k},\\
 \tilde G_{\bar mk}(\Omega)
 &=& \displaystyle \sum_{\bar m^{\prime}} R_{\bar m\bar m^{\prime}}(\Omega) G_{\bar m^{\prime} k}.
 \eeqn\esub
 For an arbitrary one-body operator $\hat O$, such as the multipole moment operator
 $\hat T_{\lambda\mu}$, the corresponding overlap is determined by
 the mixed density,
 \beqn
 {}_a\langle 0\vert \hat T_{\lambda\mu} \vert \Omega\rangle_b
 &=& \sum_{mm^\prime}(T_{\lambda\mu})_{mm^\prime} \rho^{++}_{m^\prime m}(q_a,q_b;\Omega)\nonumber\\
 & & +\sum_{\bar m\bar m^\prime}(T_{\lambda\mu})_{\bar m\bar m^\prime}
     \rho^{--}_{\bar m^\prime \bar m}(q_a,q_b;\Omega).
 \eeqn

Finally we obtain for the mixed densities in coordinate space
\beqn%
\label{mixedr}%
\rho(\bm{r};q_a,q_b;\Omega)
 &=&\sum_{mm^\prime} \rho^{++}_{mm^\prime}
       \langle\Phi_{m^\prime}(\bm{r})|\Phi_{m}(\bm{r})\rangle\\
 && \pm\sum_{\bar m\bar m^\prime} \rho^{--}_{\bar m\bar m^\prime}
       \langle\Phi_{\bar m^\prime}(\bm{r})|\Phi_{\bar
       m}(\bm{r})\rangle,\nonumber
 \eeqn%
 where the lower sign holds for the scalar density $\rho_S$ in Eq.~(\ref{E13a})
 and the upper sign for the vector density $\rho_V$ in
 Eq.~(\ref{E13b}).
 The rotation operator $\hat R(\Omega)$ does not commute with the
 reflections on the $x=0$, $y=0$, and $z=0$ planes.
 Therefore one has to extend the coordinate
 representation of the mixed density $\rho(\bm{r};q_a,q_b;\Omega)$ from 1/8 to 1/2 of
 the full space, leaving only parity and isospin projection as good quantum
 numbers.

Considering the fact that the time reversal operation $\hat {T}$
 commutes with spatial rotations $\hat R(\Omega)$
 and time reversal invariance of the quasiparticle vacua:
 $\hat {T}\vert0\rangle_a=\vert0\rangle_a,
 \hat {T}\vert\Omega\rangle_b=\vert\Omega\rangle_b$,
 one finds that the contributions from spin up and down to the mixed density $\rho(\bm{r};q_a,q_b;\Omega)$ are
 complex conjugate to each other,
 \beq
  \rho(\bm{r},\sigma;q_a,q_b;\Omega)
 = \rho^\ast(\bm{r},-\sigma;q_a,q_b;\Omega),
 \eeq
 where the relation
 \beq \hat {T}^{-1}a^\dagger_{\bm{r},\sigma}\hat {T}=-2\sigma
 a^\dagger_{\bm{r},-\sigma}
 \eeq
 has been used. This shows that the mixed densities
 $ \rho(\bm{r};q_a,q_b;\Omega)$ in coordinate space,
 summed over the spin index $\sigma$ are real.

 Moreover, there are non-vanishing mixed currents $\bj(\bm{r};q_a,q_b;\Omega)$
 with matrix elements of the same form as the densities.
 \beqn%
 \label{mixedj}
 \bj (\bm{r};q_a,q_b;\Omega)&=& -i\sum_{m {\bar m}^\prime}
 \rho^{+-}_{m{\bar m}^\prime} \langle\Phi_{{\bar m}^\prime}(\bm{r})|\bsig|\Phi_{m}(\bm{r})\rangle
~~~~~~~~~~\nonumber\\
 &&+i\sum_{{\bar m}m^\prime}
 \rho^{-+}_{{\bar m}m^\prime}
  \langle\Phi_{m^\prime}(\bm{r})|\bsig|\Phi_{\bar m}(\bm{r})\rangle.
 \eeqn%
 Since the total wave functions $\vert\Phi(\bm{r},q)\rangle$ are invariant
 under time reversal, these real part of these currents vanishes.

The mixed kinetic energy in Eq.~(\ref{Hkin}) is given by
 \beqn%
\label{B25}%
\tau(\bm{r};q_a,q_b;\Omega)
  &=&- \sum_{m {\bar m}^\prime}\rho^{+-}_{m{\bar m}^\prime}
  \langle\Phi_{{\bar m}^\prime}(\bm{r})|\bsig\cdot\bm{\nabla}|\Phi_{m}(\bm{r})\rangle
~~~~~~~\nonumber\\%
 &+&\sum_{{\bar m}m^\prime} \rho^{-+}_{{\bar m}m^\prime}
 \langle\Phi_{m^\prime}(\bm{r})|\bsig\cdot\bm{\nabla}|\Phi_{\bar m}(\bm{r})\rangle\\%
 &-&m[\rho_V(\bm{r};q_a,q_b;\Omega)-\rho_S(\bm{r};q_a,q_b;\Omega)].\nonumber%
\eeqn

Using time reversal invariance and
\bsub%
 \beqn
 {}_a\langle0\vert a_{\bar k}a_{k} \vert\Omega\rangle_b
   &=&-{}_a\langle0\vert a_{k}a_{\bar k} \vert\Omega\rangle^\ast_b,\\
   {}_a\langle0\vert a^\dagger_{k}a^\dagger_{\bar k}
 \vert\Omega\rangle^\ast_b
   &=&-{}_a\langle0\vert a^\dagger_{\bar k}a^\dagger_{k}
   \vert\Omega\rangle^\ast_b,
\eeqn%
\esub%
we obtain for the mixed pairing tensor in Dirac-space 
\bsub%
\label{C13}
\beqn%
\kappa^{10}_{k{\bar k}}
  &=&\left[{\bar R}(\Omega)\sqrt{f^b}{\bar V}^{b\ast}{\bar D}^{-1}
  {\bar U}^{aT}\sqrt{f^a}\right]^{}_{k{\bar k}},~~~~~~~~~~~~\\%
\kappa^{01}_{k{\bar k}}
  &=&\left[{\bar R}^\ast(\Omega)\sqrt{f^b}{\bar U}^{b\ast}{\bar D}^{-1}
  {\bar V}^{aT}\sqrt{f^a}\right]^\ast_{k{\bar k}},%
\eeqn%
\esub%
and we find in analogy to Eq.~(\ref{rho++}) for the mixed pairing
tensors in oscillator space
\bsub%
\label{kappa++}
\beqn%
\kappa^{10++}_{m{\bar m}^\prime}
  &=&\left[\tilde F^b(\Omega)\sqrt{f^b}{\bar V}^{b\ast}{\bar D}^{-1}
  {\bar U}^{aT}\sqrt{f^a}F^{aT}\right]^{}_{m{\bar m}^\prime},~~~~~~~~~~~~\\%
\kappa^{10--}_{{\bar m}m^\prime}
  &=&\left[\tilde G^b(\Omega)\sqrt{f^b}{\bar V}^{b\ast}{\bar D}^{-1}
  {\bar U}^{aT}\sqrt{f^a}G^{aT}\right]^{}_{{\bar m}m^\prime},~~~~~~~~~~~~\\%
\kappa^{01++}_{m{\bar m}^\prime}
  &=&\left[\tilde F^{b\ast}(\Omega)\sqrt{f^b}{\bar U}^{b\ast}{\bar D}^{-1}
  {\bar V}^{aT}\sqrt{f^a}F^{aT}\right]^\ast_{m{\bar m}^\prime},~~~~~~\\%
\kappa^{01--}_{{\bar m}m^\prime}
  &=&\left[\tilde G^{b\ast}(\Omega)\sqrt{f^b}{\bar U}^{b\ast}{\bar D}^{-1}
  {\bar V}^{aT}\sqrt{f^a}G^{aT}\right]^\ast_{{\bar m}m^\prime},~~~~~~%
\eeqn%
\esub%
and in coordinate space
\bsub%
\label{mixedk}%
\beqn%
\kappa^{10}(\bm{r};q_a,q_b;\Omega)
 &=&\sum_{m{\bar m}^\prime,\sigma} \kappa^{10++}_{m{\bar m}^\prime}
       \Phi_{{\bar m}^\prime}(\bm{r},\sigma)\Phi_{m}(\bm{r},\sigma)~~~~~~~~~~~\\
 &+& \sum_{{\bar m}m^\prime,\sigma} \kappa^{10--}_{{\bar m}m^\prime}
       \Phi_{m\prime}(\bm{r},\sigma)\Phi^{}_{\bar m}(\bm{r},\sigma),~~~~~~~~~~~\\
\kappa^{01}(\bm{r};q_a,q_b;\Omega)
 &=&\sum_{m{\bar m}^\prime,\sigma} \kappa^{01++}_{m{\bar m}^\prime}
       \Phi^\ast_{{\bar m}^\prime}(\bm{r},\sigma)\Phi^\ast_{m}(\bm{r},\sigma)\\%
 &+& \sum_{{\bar m}m^\prime,\sigma} \kappa^{01--}_{{\bar m}m^\prime}
       \Phi^\ast_{m^\prime}(\bm{r},\sigma)\Phi^\ast_{\bar m}(\bm{r},\sigma)
 \eeqn%
\esub%

 In this investigation, GCM and configuration mixing is not taken into account.
 Therefore we have $\vert 0\rangle_a=\vert 0\rangle_b$ and only diagonal contractions
 with $q_a=q_b=q$.

 \section{Symmetries in overlaps}

 \label{AppendixC}
 \subsection{Symmetries associated with $\phi$ and $\psi$}
 The $D_2$ symmetry and time reversal symmetry have been imposed in
 the mean-field calculation, which leads to the mean-field state $\vert \Phi(q)\rangle$
 invariant under the following transformations,
 \beqn
 e^{i\pi\hat J_k}\vert \Phi(q)\rangle=\vert \Phi(q)\rangle,
 \quad k=x,y,z.
 \eeqn
 It reduces the integration intervals for the Euler angles $(\phi,\theta,\psi)$
 in Eqs.~(\ref{Integration1}) and (\ref{Integration2})
 to $\phi\in[0,\pi]$, $\theta\in[0, \pi]$, $\psi\in[0,\pi]$.
 The Hamiltonian kernel $H^J_{KK^\prime}$ and the norm kernel
 $N^J_{KK^\prime}$ are simplified as
 \beqn
  O^J_{KK^\prime}(q,q)
  &=&{\cal F}_{KK^\prime}\frac{2J+1}{8\pi^2}
    \int^{\pi}_0d\phi\int^{\pi}_0d\theta \int^{\pi}_0d\psi\nonumber\\
  &&\times  \langle \hat O \hat R(\phi,\theta,\psi)\rangle
      D^{J\ast}_{KK^\prime}(\phi,\theta,\psi),
  \eeqn
 where $O=1, \hat H$ and the factor
 ${\cal F}_{KK^\prime}=1 + e^{-iK\pi}+ e^{-iK^\prime\pi}+ e^{-i(K+K^\prime)\pi}$.
 Furthermore, the rotation operator $\hat R(\phi,\theta,\psi)$ is
 transformed as
 \beq
   e^{-i\pi\hat J_x}\hat R(\phi,\theta,\psi)e^{i\pi\hat J_x}
  =\hat R(-\phi,-\theta,-\psi).
 \eeq
 The many-body Hamiltonian $\hat H$ is rotational invariant, which
 leads to together with orthogonality to the following symmetry
 relations for the Hamiltonian overlap
 \beqn
  \langle \hat H\hat R(\phi,\theta,\psi)\rangle
   &=&\langle  \hat H\hat R(-\phi,-\theta,-\psi)\rangle,\\
  \langle \hat H\hat R(\phi,\theta,\psi)\rangle^\ast
  &=&\langle \hat H\hat R(-\psi,-\theta,-\phi)\rangle,\\
  \langle \hat H\hat R(\phi,\theta,\psi)\rangle^\ast
  &=&\langle \hat H\hat R(\psi,\theta,\phi)\rangle.
  \eeqn
 With the help of relation:
  $e^{i\pi\hat J_z}e^{-i\theta\hat J_y}e^{-i\pi\hat J_z}
  =e^{i\theta\hat J_y}$, one gets
  \beqn
  \langle \hat H\hat R(\phi,\theta,\psi)\rangle^\ast
  &=&\langle \hat H\hat R(-\psi,-\theta,-\phi)\rangle\nonumber\\
  &=&\langle \hat H\hat R(\pi-\psi,\theta,\pi-\phi)\rangle\nonumber\\
  &=&\langle \hat H\hat R(\pi-\phi,\theta,\pi-\psi)\rangle,
  \eeqn
  which can also be derived from the reality condition:
  \beqn
  \langle \hat H\hat R(\phi,\theta,\psi)\rangle^\ast
  &=&\langle \hat H\hat R(-\phi,\theta,-\psi)\rangle\nonumber\\
  &=&\langle \hat H\hat R(\pi-\phi,\theta,\pi-\psi)\rangle.
  \eeqn

 In a similar way we can derive symmetries of the overlaps with
 $\langle\hat T_{\lambda\mu}\hat{R}(\Omega)\rangle$. Since $\hat T_{\lambda\mu}$
 is not rotational invariant, the overlaps with the Euler angles $\phi, \psi$
 in regions $[0,\pi]$ and $[\pi,2\pi]$ are related by the following relations,
 \bsub%
 \beqn%
 \langle \hat T_{\lambda\mu}\hat R(\pi+\phi,\theta,\psi)\rangle
  &=&(-1)^\mu\langle \hat T_{\lambda\mu}\hat R(\phi,\theta,\psi)\rangle,\\
  \langle \hat T_{\lambda\mu}\hat R(\phi,\theta,\pi+\psi)\rangle
  &=& \langle \hat T_{\lambda\mu}\hat R(\phi,\theta,\psi)\rangle.
 \eeqn%
 \esub%
 The tensor $\hat T_{\lambda\mu}$ is transformed under $e^{-i\pi\hat J_x}$ as,
 \beq
  e^{-i\pi\hat J_x}\hat T_{\lambda\mu}  e^{i\pi\hat J_x}
  =(-1)^\lambda \hat T_{\lambda-\mu},
 \eeq
 which gives rise to the symmetry:
 \beq
 \langle \hat T_{\lambda\mu}\hat R(\phi,\theta,\psi)\rangle
 =(-1)^\lambda
 \langle \hat T_{\lambda-\mu}\hat R(\pi-\phi,\theta,\pi-\psi)\rangle.
 \eeq

 \subsection{Symmetries associated with $\theta$}
 Since the mean-field state $\vert \Phi(q)\rangle$ is invariant
 under the transformation $ e^{i\pi\hat J_y}$,
 \beqn
 \langle \hat H\hat R(\phi,\pi-\theta,\psi)\rangle
 &=&\langle \hat He^{i\phi\hat J_z}e^{-i\theta\hat J_y}
    e^{i\pi\hat J_y}e^{i\psi\hat J_z} \rangle\nonumber\\
 &=&\langle \hat H\hat R(\phi,-\theta,-\psi)\rangle\nonumber\\
 &=& \langle \hat H\hat R(\phi,\theta,\pi-\psi)\rangle^\ast.
 \eeqn
 On the other hand, the group elements in the group $D_2$ obey the relation:
 $e^{i\pi\hat J_y}=e^{i\pi\hat J_x}e^{i\pi\hat J_z}$,
 \beqn
 \langle \hat H\hat R(\phi,\pi-\theta,\psi)\rangle
 &=&\langle \hat H
    e^{i\phi\hat J_z}e^{i\pi\hat J_x}e^{i\pi\hat J_z}
    e^{-i\theta\hat J_y} e^{i\psi\hat J_z} \rangle\nonumber\\
 &=&\langle \hat H\hat R(\phi,\theta,-\psi)\rangle\nonumber\\
 &=& \langle \hat H\hat R(\phi,\theta,\pi-\psi)\rangle.
 \eeqn
This shows that the Hamiltonian overlap is real.
 With the help of the relation: $e^{i\pi\hat J_z}\hat T_{\lambda\mu} e^{-i\pi\hat J_z}
  =(-1)^\mu \hat T_{\lambda\mu}$, one finds the symmetry,
 \beq
 \langle \hat T_{\lambda\mu}R(\phi,\pi-\theta,\psi)\rangle
 =(-1)^\mu
  \langle \hat T_{\lambda\mu}R(\phi,\theta,\pi-\psi) \rangle.
 \eeq
 These symmetries of the hamiltonian overlap integrals simplify the calculations
 considerably by reducing the necessary interval, where the overlap integrals have
 to be calculated from $[0,\pi]$ to $[0,\pi/2]$.

 \end{appendix}


\end{document}